\documentclass[fleqn,usenatbib]{mnras}

\usepackage{newtxtext,newtxmath}

\usepackage[T1]{fontenc}

\DeclareRobustCommand{\VAN}[3]{#2}
\let\VANthebibliography\thebibliography
\def\thebibliography{\DeclareRobustCommand{\VAN}[3]{##3}\VANthebibliography}

\usepackage{graphicx}	
\usepackage{amsmath}	
\usepackage{chemformula}
\usepackage{xspace}
\usepackage{upgreek}
\usepackage[version=4]{mhchem} 
\usepackage{natbib}
\usepackage{orcidlink}
\usepackage{subfig}
\usepackage{pdflscape}
\usepackage[flushleft]{threeparttable}
\usepackage{lineno}

\newcommand{\tiberius}[1]{\texttt{Tiberius}\xspace}
\newcommand{\jedi}[1]{\texttt{ExoTiC-JEDI}\xspace}
\newcommand{\poseidon}[1]{\texttt{POSEIDON}\xspace}
\newcommand{\prt}{\texttt{petitRADTRANS}\xspace}

\newcommand\micro{$\upmu$}
\newcommand\um{\micro m\xspace}
\defcitealias{maymacdonald2023g395h_gj1132b}{May \& MacDonald et al. 2023}

\title[BOWIE-ALIGN: NGTS-2b]{BOWIE-ALIGN: Exploring degeneracies in the muted transmission spectrum of the aligned hot Jupiter NGTS-2b with NIRSpec/G395H.}

\author[C. Fairman et al.]{Charlotte Fairman,$^{\orcidlink{0000-0001-9665-5260}~1}$\thanks{E-mail: charlotte.fairman@bristol.ac.uk},
Hannah R. Wakeford,$^{\orcidlink{0000-0003-4328-3867}~1}$,
Alastair B. Claringbold$^{2,3}$, 
James Kirk$^{\orcidlink{0000-0002-4207-6615}~4}$, \newauthor
Eva-Maria Ahrer$^{\orcidlink{0000-0003-0973-8426}~5}$,
Daniel Thorngren$^{\orcidlink{0000-0002-5113-8558}~6}$, 
Shang-Min Tsai$^{\orcidlink{0000-0002-8163-4608}~7}$,
R. A. Booth$^{\orcidlink{0000-0002-0364-937X}~8}$,
Anna B. T. Penzlin$^{\orcidlink{0000-0002-8873-6826}9}$, \newauthor
Lili Alderson$^{\orcidlink{0000-0001-8703-7751}~10}$,
Duncan A. Christie$^{\orcidlink{0000-0002-4997-0847}~5}$,
M. L\'opez-Morales$^{\orcidlink{0000-0003-3204-8183}~11}$,
N. J. Mayne$^{\orcidlink{0000-0001-6707-4563}~12}$, 
Annabella Meech$^{\orcidlink{0000-0002-7500-7173}~13}$, \newauthor 
James E. Owen$^{\orcidlink{0000-0002-4856-7837}~4}$,
Vatsal Panwar$^{\orcidlink{0000-0002-2513-4465}~14}$,
Daniel Valentine$^{\orcidlink{0000-0002-2643-6836}~1}$,
Peter J.\ Wheatley$^{\orcidlink{0000-0003-1452-2240}~2,3}$,
Maria Zamyatina$^{\orcidlink{0000-0002-9705-0535}~12}$\newauthor
\\
$^{1}$University of Bristol, School of Physics, HH Wills Physics Laboratory, Tyndall Avenue, Bristol, BS8 1TL\\
$^{2}$Centre for Exoplanets and Habitability, University of Warwick, Gibbet Hill Road, Coventry CV4 7AL, UK\\
$^{3}$Department of Physics, University of Warwick, Gibbet Hill Road, Coventry CV4 7AL, UK\\
$^{4}$Department of Physics, Imperial College London, Prince Consort Road, SW7 2AZ, London, UK\\
$^{5}$Max Planck Institute for Astronomy (MPIA), K\"{o}nigstuhl 17, 69117 Heidelberg, Germany \\
$^{6}$William H. Miller III Department of Physics \& Astronomy, Johns Hopkins University, 3400 N Charles St, Baltimore, MD 21218, USA \\
$^{7}$ Institute of Astronomy \& Astrophysics, Academia Sinica, Taipei 10617, Taiwan\\
$^{8}$School of Physics and Astronomy, University of Leeds, Leeds LS2 9JT, UK\\
$^{9}$Ludwig-Maximilians-Universit{\"a}t M{\"u}nchen, Universit{\"a}ts-Sternwarte, Scheinerstr.~1, 81679 M{\"u}nchen, Germany\\
$^{10}$Department of Astronomy, Cornell University, 122 Sciences Drive, Ithaca, NY 14853, USA\\
$^{11}$Space Telescope Science Institute, 3700 San Martin Drive, Baltimore MD 21218, USA\\
$^{12}$Department of Physics and Astronomy, Faculty of Environment, Science and Economy, University of Exeter, Exeter EX4 4QL, UK\\
$^{13}$Center for Astrophysics | Harvard \& Smithsonian, 60 Garden St, Cambridge, MA 02138, USA\\
$^{14}$School of Physics and Astronomy, University of Birmingham, Edgbaston, Birmingham B15 2TT, UK\\
}

\date{Accepted XXX. Received YYY; in original form ZZZ}

\pubyear{\the\year{}}

\begin{document}
\label{firstpage}
\pagerange{\pageref{firstpage}--\pageref{lastpage}}
\maketitle

\begin{abstract}
We present the first atmospheric observation and characterisation of the aligned, 1468\,K hot Jupiter, NGTS-2b, with one JWST NIRSpec/G395H transit. These observations complete the GO 3838 observing campaign of the BOWIE-ALIGN program, which aims to investigate the link between hot Jupiter atmospheric composition and formation history through the atmospheric analysis of planets orbiting F stars that are aligned and misaligned with the host stellar spin axis. The 2.84--5.18\,\um spectrum shows weak absorption features attributed to \ce{H2O} and \ce{CO2} absorption, which our free chemistry retrievals fit with posteriors that converge on high mean molecular weight solutions attained through significant \ce{H2O} mixing ratios. By comparing our results to interior modelling, we show that some of these solutions exceed the $43.5\times$ solar upper limit we obtained from our interior structure models. Such solutions are likely due to cloud-metallicity degeneracies and insufficient wavelength coverage to resolve them. We show that, in the case of our observations, the likelihood distribution of \ce{H2O} abundances is flat and uninformative, such that our retrievals are biased by the prior. Additionally, our statistically favoured atmospheric solution contains absorption from \ce{SO}. The chemical abundances retrieved with this model are likely not astrophysically feasible and we demonstrate that the presence of \ce{SO} is driven by only two data points. Our equilibrium chemistry retrievals hint at a subsolar C/O ratio and supersolar metallicity; however, we find wide posterior distributions that extend to solar values.
\end{abstract}




\section{Introduction}
An open question in the study of exoplanet atmospheres is how observations of present-day atmospheric composition can be used to infer formation history. Hot Jupiters, due to their large radii and strong irradiation from their host stars, provide the best opportunity for robust atmospheric characterisation through transmission spectroscopy \citep[e.g.,][]{Sing2016,Kempton2017_exotransmit,Kempton2018_TMM}. It is unlikely that such planets could have formed in situ through either gravitational instability or core accretion \citep[e.g.,][]{Dawson2018}, so they are expected to have migrated through the disc to their present day location. As such, their formation location and migration history should impart signatures within their bulk and atmospheric composition \citep[e.g.,][]{Oberg2011, madhusudhan2014, Schneider_Bitsch2021,Penzlin2024_BOWIE}.

\renewcommand{\arraystretch}{1.2}
\begin{table*}
    \begin{threeparttable}
    \centering
    \caption{Fitted system parameters of NGTS-2b from the NIRSpec/G395H NRS1 and NRS2 white light curves. We present the results from two independent reductions with \jedi{} and \tiberius{}. }
    \label{tab:fitted_system_parameters}
    \begin{tabular}{l c c c c c} 
    \hline
     & & $R_p/R_*$ & $a/R_*$ & $T_0$ (BJD)  & $i$ ($^\circ$)  \\
     \hline 
     Literature Value &  & $0.09619^{+0.00114}_{-0.00088}$$^*$ & $8.0\pm0.4$$^\dag$  & - & $88.5^{+1.0}_{-1.2}$$^\dag$   \\
     \hline
     \jedi & & NRS1 & 0.099300 $\pm$ 0.000064 & 7.718 $\pm$ 0.029 &  60501.389230 $\pm$ 0.000035 & 87.321 $\pm$ 0.079 \\
           & NRS2 & 0.098894 $\pm$ 0.000077 & 7.708 $\pm$ 0.036 & 60501.389199 $\pm$ 0.000043 & 87.290 $\pm$ 0.096 \\
           & Weighted Mean & 0.099136 $\pm$ 0.000070 & 7.714 $\pm$ 0.033 & 60501.389218 $\pm$ 0.000039 & 87.308 $\pm$ 0.088 \\
     \hline
     \tiberius & & NRS1 & 0.099446 $\pm$ 0.000059 & 7.721 $\pm$ 0.028 & 60501.389237 $\pm$ 0.000033 & 87.311 $\pm$ 0.075 \\
           & NRS2 & 0.099030 $\pm$ 0.000077 & 7.689 $\pm$ 0.036 & 60501.389218 $\pm$ 0.000043 & 87.220 $\pm$ 0.095 \\
           & Weighted Mean & 0.099292 $\pm$ 0.000047 & 7.709 $\pm$ 0.022 & 60501.389230 $\pm$ 0.000026 & 87.276 $\pm$ 0.059 \\ 
     \hline

    \end{tabular}
    \begin{tablenotes}
        \item[*] Values from \citet{raynard2018ngts2b}
        \item[$\dag$] Values from \citet{kokori2023}
    \end{tablenotes}
    \end{threeparttable}
\label{table:retrieval-results-species}
\end{table*}

At first-order, a planet’s formation location relative to ice lines of oxygen- and carbon-rich species in its protoplanetary disc will affect the C/O and C/H ratios of its gaseous envelope \citep[e.g.][]{Oberg2011}. However, there are higher order factors that introduce additional complexity to this picture, making the direct inference of formation history from bulk envelope composition challenging \citep[e.g.,][]{notsu2020, Molliere2022,Feinstein2025_formation}. Compositional changes over a protoplanetary disc's lifetime will influence planetary C/O ratio, such as through the evolution of ice lines \citep{morbidelli2016,Eistrup2018,Owen2020}, inward pebble drift \citep{oberg_bergin2016,Booth2017,Bosman2018}, and gas-phase and grain-surface reactions \citep{Eistrup2018,Molliere2022}. 
Similarly, additional aspects of planet evolution can contribute to the final composition of the planetary atmosphere such as hydrodynamic escape and atmospheric mass loss \citep[e.g.,][]{Malsky2023}; however, hot Jupiters are massive enough that escape cannot alter their primordial composition \citep[e.g.,][]{Owen_Wu2016,Owen_Adams2016}.

On inferring the composition of exoplanetary atmospheres through transmission spectroscopy, an additional uncertainty in linking such measurements to planet formation lies in whether the underlying atmospheric composition is representative of bulk composition \citep[e.g.][]{Feinstein2025_formation}. Interior modelling of giant exoplanets often assumes exoplanets to host fully convective interiors, leading to well mixed atmospheres \citep[e.g.,][]{Miller_fortney2011,Miguel2016,Thorngren2016}. However, there is evidence that giant exoplanets may contain deep radiative zones which can inhibit interior-atmosphere mixing, leading to stably stratified atmospheres and compositional gradients \citep{KnierimHelled2024ApJ}.

Due to the complexity of planet formation processes, directly inferring formation history from the atmospheric composition of a single planet is not predictive. However, measuring relative trends between populations is less challenging and provides a more testable hypothesis of planet formation \citep{Kirk2024}. \citep{Penzlin2024_BOWIE} showed the relative differences in atmospheric composition between populations of aligned and misaligned planets can be distinguishable, despite these complexities.
This framework motivates the BOWIE-ALIGN survey (GO-3838, PIs Kirk \& Ahrer) outlined in \citet{Kirk2024}, which aims to analyse a sample of eight hot Jupiters, half of which are aligned with their host stellar spin axis (|$~\lambda$~|$<$ 30$^\circ$), whilst the other four are misaligned (|$~\lambda$~|$>$ 45$^\circ$). The underlying hypothesis is that aligned planets have migrated through the protoplanetary disc, accreting material throughout the disc as the planet’s orbit shrinks \citep[e.g.,][]{Dawson2016}.  Conversely, misaligned planets are thought to have formed further from their host star and reached their current location via high eccentricity migration after disc dispersal \citep[e.g.,][]{wu_murray2003,munoz2016}. The sample specifically contains hot Jupiters orbiting F stars above the Kraft break ($T_\mathrm{eff} \gtrapprox$ 6200\,K, \citealt{Kraft1967}), marking the effective temperature boundary below which stellar rotation rates dramatically decrease \citep[e.g.,][]{winn2010, albrecht2012}. 
This means that the sample of planets in BOWIE-ALIGN should not be strongly impacted by tidal realignment as the forces are smaller for these stars \citep{dawson2014}, making the migration history between aligned and misaligned targets correspond to distinct migration scenarios, and thus differing C/O and C/H ratios.

To date, the BOWIE-ALIGN program has presented the transmission spectrum of four planets, two aligned and two misaligned. A super-solar metallicity and C/O ratio consistent with solar was found for the misaligned planet, WASP-15b \citep{kirk2025_w15}, with tentative evidence for the presence of sulphur-based species. 
Aligned and misaligned respectively, TrES-4b \citep{meech2025_t4} and HAT-P-30b \citep{claringbold26h30} were both found to host sub-solar metallicities and C/O ratios, a composition more easily explained through oxygen-rich gas accretion.  
Finally, the transmission spectrum of the aligned planet, KELT-7b \citep{ahrer2025_k7}, showed muted features, suggesting either a metal depleted atmosphere or the presence of clouds. Solar to supersolar metallicities were preferred for KELT-7b from equilibrium retrievals; however, C/O ratios were poorly constrained, making inferences to potential formation histories challenging.  In this work, we complete the GO 3838 observing campaign with the JWST NIRSpec/G395H transmission spectrum of NGTS-2b. In future work (Ahrer et. al. in prep., Kirk et. al. in prep.), we will fold in the archival observations of the misaligned WASP-94Ab \citep{ahrer2025wasp94}  and WASP-17b (Lewis et. al. in prep.), and the aligned HD 149026b \citep{bean2023} to complete the eight planet BOWIE-ALIGN sample.

NGTS-2b is a $0.74^{+0.13}_{-0.12}\, M_\mathrm{J}$, $1.595^{+0.047}_{-0.045}\, R_\mathrm{J}$ planet with an equilibrium temperature of $1468^{+45}_{-42}$ K \citep{raynard2018ngts2b}, independently discovered by both \citet{raynard2018ngts2b} and \citet[][as WASP-179b]{Anderson2018arXiv_W179}. Orbiting an F5V host star ($1.64^{+0.19}_{-0.22}\, M_\odot$; $1.702^{+0.047}_{-0.044}\, R_\odot$, $T_\mathrm{eff} = 6478^{+94}_{-89}$ K, \citealt{raynard2018ngts2b}), NGTS-2b falls into the aligned sample of planets, with a measured stellar obliquity through line profile tomography of $\lambda  = -11 \pm 5^\circ$ \citep{Anderson2018arXiv_W179}. Here we present the first spectroscopic observations and atmospheric characterisation of NGTS-2b. In Section \ref{sec:data} we outline our observations and two data reduction pipelines applied to the measurements. In Section\,\ref{sec:interior} we discuss the implications of interior structure modelling for this planet. Section\,\ref{sec:retrievals} details our retrieval frameworks and their results, with the median retrieved values and full posterior distributions presented in Appendix \ref{app:retrieval_results} and Appendix \ref{sec:cornerplots} respectively. We discuss the implications of our results on the atmosphere of NGTS-2b in Section\,\ref{sec:discussion}. Finally, in Section\,\ref{sec:conclusions} we present our conclusions from this study.

\section{Observation and Data Reduction}\label{sec:data}

We observed one transit of NGTS-2b with the JWST NIRSpec/G395H grating on the 10$\mathrm{th}$ July 2024 between 03:03:47 and 13:17:08 UT.
This observation provides spectroscopy between 2.84--5.14$\upmu$m (with a gap between the NRS1 and NRS2 detectors, spanning 3.72--3.82$\upmu$m) at an average spectral resolution of $R\sim 2700$. We carried out the observation using the Bright Object Time Series (BOTS) mode, using the F290LP filter, SUB2048 subarray, and NRSRAPID readout pattern. The observation lasted 10.22 hours, with a total of 706 integrations and 48 groups per integration.  

We used two data reduction pipelines (\jedi{} and \tiberius{}) to assess our data quality and test how robust the resultant planetary transmission spectrum was to decisions made in the analysis of the data. We detail each of these processes below and discuss their comparative spectra in section \ref{sec:spec}.

\subsection{ExoTiC-JEDI reduction}

To produce integration images from our raw \texttt{uncal} files, we employ Stage 1 of the \jedi{} pipeline \citep{alderson2022jedi}, which wraps features of the \texttt{jwst} pipeline (v1.15.1) with custom functions and has been used in a number of previous datasets (e.g., \citetalias{maymacdonald2023g395h_gj1132b}; \citealt{ alderson2023g395h_w39ers, alderson2024compass, scarsdale2024compass,ahrer2025_k7}). We first apply a custom bias subtraction over the default \texttt{jwst} pipeline bias correction. This generates a pseudo bias image, representing the median pixel value of the first group across all integrations. This image is then subtracted from all groups in the observation. Default \texttt{jwst} routines are implemented to perform the linearity, dark current, and saturation corrections. To identify any persisting nonlinearity in the group level ramps due to offsets between subsequent groups, we apply a jump step threshold of 15$\sigma$. 

Before fitting the ramps, we perform a custom group level $1/f$ noise correction. To isolate the trace, a mask spanning 15 $\sigma$ from the expected trace position is created. For each cross-dispersion column in a group image, the median of the remaining background region is evaluated and subtracted. A standard linear ramp is then fit to each integration using least-squares minimization, from which the mean count rate for each integration is evaluated. We obtain the DQ flag array and 2D wavelength map from the \texttt{jwst} pipeline for use in the data cleaning and light curve extraction of \jedi{} Stage 2, described below.

An initial trimming of the data along the dispersion direction is performed to isolate the trace on each detector; the first 500 and 5 columns are removed from NRS1 and NRS2, respectively. 
We replace pixels flagged as do not use, dead, hot, low quantum efficiency and no gain value by the \texttt{jwst} pipeline with the median of a 4-pixel window on either side along the dispersion direction. Pixels that are not flagged by the DQ array or picked up in later cleaning that we identify as contaminants are added to the DQ array for replacement. For NRS1, we flag a single pixel at (1500, 27) and for NRS2, three hot pixels with no DQ flag ((87, 15), (845,17), (1203,21)) are identified and replaced during cleaning. However, residual contamination within the surrounding pixels still remains, so we mask these pixels along with their immediate vertical and horizontal neighbours. 

\begin{figure*}
    \centering
    \includegraphics[width=\linewidth]{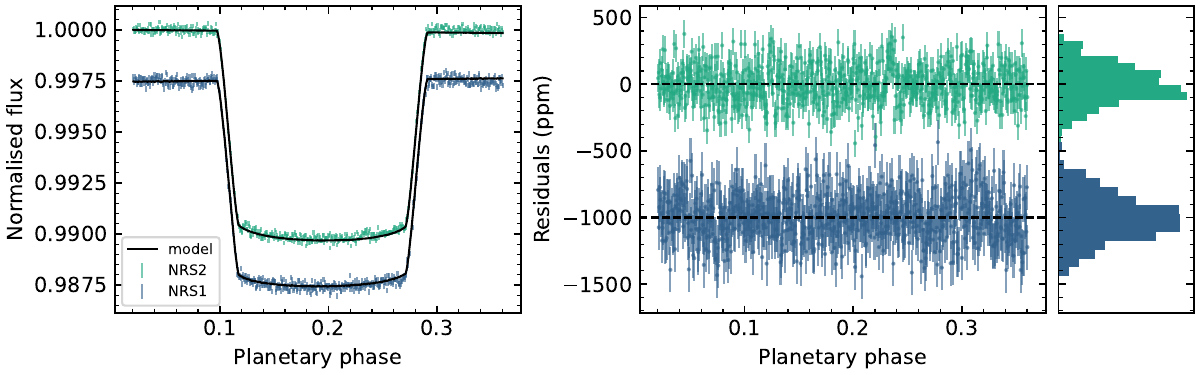}
    \caption{NGTS-2b broadband transit light curves (left) and residuals (right) from NRS1 and NRS2 using the \jedi{} reduction. Data from the different detectors are offset for clarity.}
    \label{fig:lcs}
\end{figure*}

\begin{figure}
    \centering
    \includegraphics[width=\linewidth]{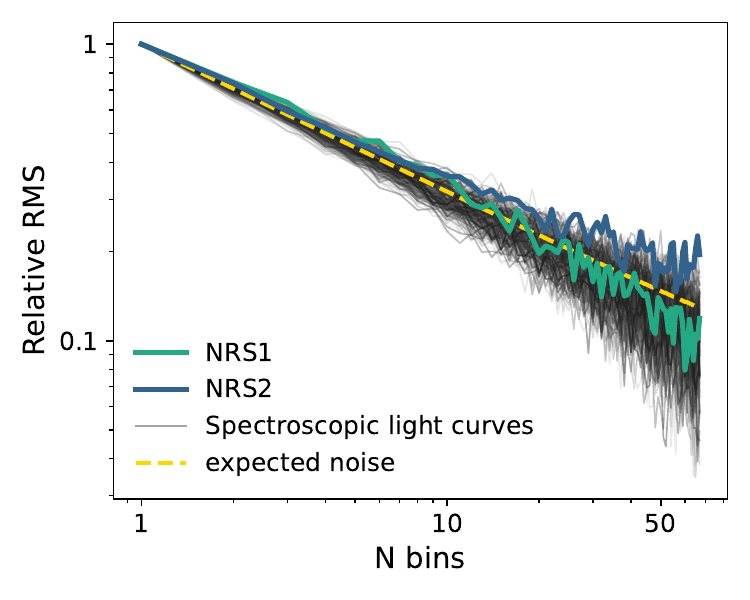}
    \caption{Binned residual plot for our ExoTiC-JEDI reduction across all spectroscopic light curves (grey) and broad band light curves for NRS1 (green) and NRS2 (blue). The dashed yellow line shows the expected noise properties as the residuals are binned down. This demonstrates that our data aligns with the expectation and is not impacted by unaccounted for red noise in the data.}
    \label{fig:red_noise}
\end{figure}

Next, to identify temporal outliers, we search through the time domain with a window size of 10 pixels, identifying outlying pixels greater than an $n$-$\sigma$ threshold to replace with the median window value. 
To distinguish between the stellar signal and background noise, we optimized the sigma significance at which we clean temporal outliers by identifying where the standard deviation at which the number of outliers found by the $n$-$\sigma$ threshold deviates from a linear least squares fit by 0.05 $\sigma$, on five points spanning thresholds of 25 to 16 $\sigma$.
This led to the adoption of thresholds of 12 and 15 $\sigma$ for NRS1 and NRS2 respectively, representing 0.080\% and 0.085\% of the total pixels on each detector.  

Subsequent cleaning of spatial outliers along the dispersion direction was then performed. A fourth order polynomial was fit across a search window of 60 pixels with any pixels falling outside 6$\sigma$ flagged as outliers.
The identified pixels were replaced with the median value calculated from 4-pixel windows adjacent to each point. This polynomial fitting is repeated until no more outlying pixels are flagged. In total, 14587 pixels (0.042\%) were identified as outliers in NRS1 and 37595 pixels (0.108\%) in NRS2. 

With the cleaned data cube, we locate the centre of the spectral trace by fitting each spatial column with a Gaussian. For later spectral extraction, the aperture of the trace is optimised to 3.5 FWHM (an average of 4.96 pixels in NRS1 and 5.52 pixels in NRS2) to maximise the post transit precision of the raw light curve. The trace width is smoothed with a median filter window of 5 pixels before the trace position and trace width are fit with fourth-order polynomials to define the extraction aperture. An additional 1/f noise destriping is implemented at this stage, isolating the background with aperture buffers 7.5 and 5.5 pixels above below the upper and lower extraction boundaries for NRS1 (7 and 5 pixels for NRS2). The median of this background region is subtracted from the signal in each column. 

We perform an intrapixel extraction to obtain our 1D stellar spectra. These spectra are cross-correlated with the median unshifted spectrum to determine the x- and y-pixel positional shifts on the detector as a function of time.

We produce broadband light curves for NRS1 and NRS2 spanning wavelengths of 2.814--3.717$\upmu$m and 3.814--5.111$\upmu$m, to which we fit a two-component astrophysical and noise model. The astrophysical component utilizes the \texttt{batman} light curve package \citep{kreidberg2015batman}, where we fit for the system parameters: the ratio of the planet radius to stellar radius ($R_p/R_*$), orbital semi-major axis ($a$), inclination ($i$), and mid transit time ($t_0$). We fix the period to 4.511123 days \citep{kokori2023} and adopt zero eccentricity based on the transit fitting of \citet{raynard2018ngts2b}. Our systematic model takes the form
\begin{equation*}
    S(\lambda) = s_0 + (s_1\, x_s\lvert y_s \rvert) + (s_2 \, t),
\end{equation*}
\noindent where $x_s$ is the shift of the trace position in the dispersion direction and $\lvert y_s \rvert$ is the absolute positional shift of the trace in the cross dispersion direction. The coefficients $s_0$, $s_1$ and $s_2$ are fitted coefficients. A number of additional systematic models were tested, with limited impact on the out of transit residuals and resultant transmission spectrum.
We use a fixed 4-parameter nonlinear limb-darkening law calculated by the \texttt{ExoTiC-LD} package \citep{ grant2024exoticld_joss}. The stellar parameters $\log{g}$, [Fe/H] and $T_\mathrm{eff}$ are adopted from the nearest grid point of the Stagger 3D stellar models \citep{magic2015stellarld} to the metallicity ([Fe/H] $= -0.06$ ), effective temperature ($T_\mathrm{eff} = 6478$ K) and surface gravity ($\log{g} = 4.197$) found in \citet{raynard2018ngts2b}.

The best fit parameters are determined using (Levenberg-Marquadt) L-M least-squares minimisation \citep{scipy}. This fitting is performed twice: initially to produce error inflation values such that the best fit model has a $\chi^2_\nu$ = 1, and then again on the rescaled flux uncertainty errors. The fitted broadband light curve parameters are presented in Table \ref{tab:fitted_system_parameters} and the broadband light curves and their residuals are shown in Figure \ref{fig:lcs}. 

\begin{figure*}
    \centering
    \includegraphics[width=\linewidth]{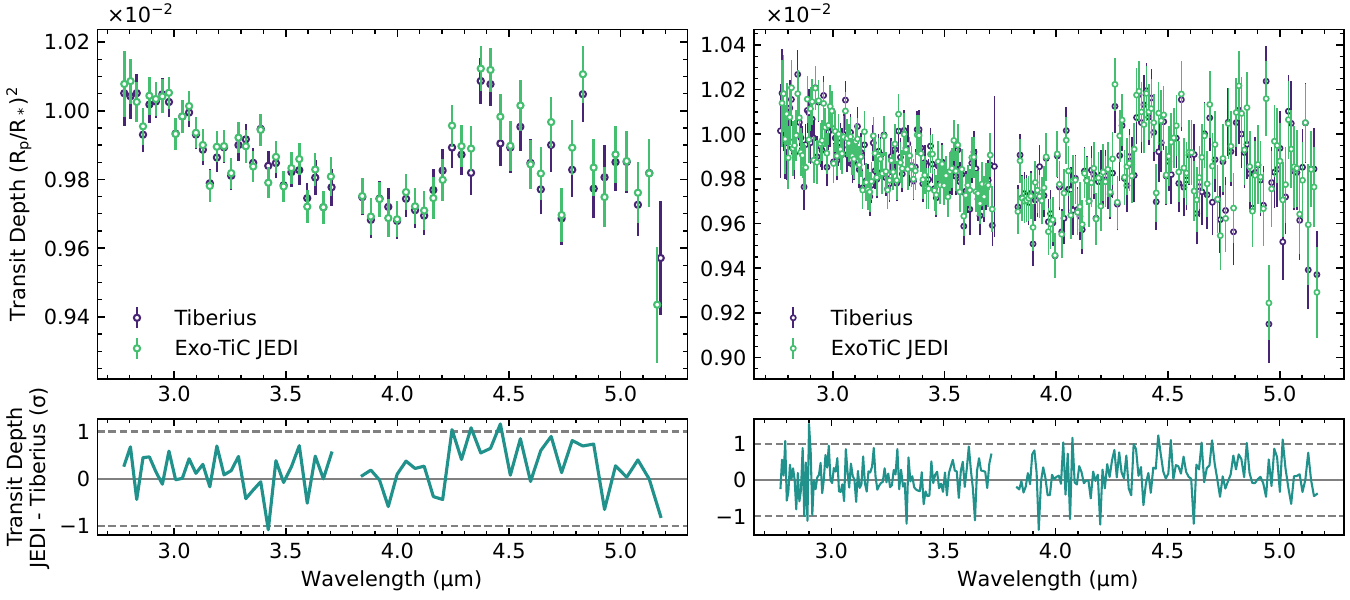}
    \caption{NGTS-2b transmission spectra from JWST NIRSpec/G395H at $R$\,=\,100 (left) and $R$\,=\,400 (right) using two different reduction pipelines (\jedi{} in green and \tiberius{} in purple). Bottom plots show the residuals between the two pipelines are in very good agreement within the 1$\sigma$ bounds, which are shown by the dashed lines. }
    \label{fig:spectra}
\end{figure*}

For the spectroscopic light curves, we bin the data at resolutions of $R$\,=\,100 and $R$\,=\,400. We fit for $R_p/R_*$ whilst fixing $a$, $i$ and $t_0$ to the broadband light curve values, using the same method and error inflation as above. Figure\,\ref{fig:red_noise} shows that each of our light curves bin down to the expected noise level, thus indicating that we do not have any remaining uncorrelated (`red') noise in our data. 
The final transmission spectra for each resolution are plotted in Figure\,\ref{fig:spectra}.

\subsection{Tiberius reduction}

For a second reduction of the data, we used the \texttt{Tiberius} data reduction package \citep{Kirk2017,Kirk2021} which has been used in a number of studies of JWST data to date \citep[e.g.,][]{Esparza-Borges2023,Kirk2024_GJ341b,Powell2024}. Our reduction of the NGTS-2b data proceeded in an identical way to that of the other BOWIE-ALIGN targets (e.g., WASP-15b \citealt{kirk2025_w15}). We did this to ensure one uniform reduction approach threading throughout all BOWIE-ALIGN analyses.

A detailed description of this process is given in \cite{kirk2025_w15}. In brief, we run stage 1 of the \texttt{jwst} pipeline (v1.8.2) on the \texttt{gainscalestep.fits} files, skip the \texttt{jump\_step}, and perform our own 1/f noise correction. We then create our own bad pixel mask and flag and replace outliers in the pixel time-series. Finally, we extract the stellar spectra between rows 608--2044 (zero indexed) for NRS1 and rows 3--2043 for NRS2, using standard aperture photometry with a fixed aperture full width of 8 pixels.

With the stellar spectra in hand, we create the spectroscopic light curves using two different binning schemes at spectral resolutions of $R$\,=\,100 and $R$\,=\,400. These are the same bins as used in our other BOWIE-ALIGN analyses.

To fit our transit light curves, we adopt the same approach as for our other BOWIE-ALIGN analyses. Specifically, our transit light curve models comprise a quadratically limb-darkened analytic transit model (implemented through \texttt{batman}, \citealt{kreidberg2015batman}) multiplied by a linear-in-time polynomial to capture systematic noise in the data. We chose this approach to be consistent our other analyses with the goal of mitigating biases when we come to compare the spectra from each of our targets. 

For the white light curves, the free parameters of our model were the planet's time of mid-transit $T_0$, inclination $i$, scaled semi-major axis $a/R_\star$, ratio of planet-to-star radii $R_P/R_\star$, and the two parameters of our linear polynomial $c_1$, $c_2$. We held the planet's orbital period fixed to 4.5111230\,d  \citep{kokori2023} and its eccentricity to 0 \citep{raynard2018ngts2b}. Similar to our other \texttt{Tiberius} analyses of BOWIE-ALIGN targets, we also fixed both quadratic limb darkening coefficients, $u_1$ and $u_2$, to values derived from \texttt{ExoTiC-LD} \citep{grant2024exoticld_joss} using the 3D Stagger grid \citep{magic2015stellarld} and the same stellar parameters as in our \jedi{} reduction.

We used a L-M algorithm implemented through \texttt{scipy} \citep{scipy} to determine the best-fitting parameters and associated $1\sigma$ uncertainties. We did this in two iterations. The first iteration was used to rescale the photometric uncertainties to give $\chi^2_{\nu} = 1$. The second iteration was used to determine our final parameter values and uncertainties. Following the white light curve fits, we fitted our spectroscopic light curves following the same approach. However, we fixed the system parameters ($T_0$, $i$ and $a/R_\star$) to the weighted mean values from the NRS1 and NRS2 white light curve fits. These values are given in Table \ref{tab:fitted_system_parameters}. Upon the completion of these fits we obtained the planet's transmission spectrum which is plotted in Figure\,\ref{fig:spectra}.

\subsection{The transmission spectrum}\label{sec:spec}
We show a comparison of our two reductions in Figure\,\ref{fig:spectra}. Overall, our results are well within one-sigma of each other, with small differences in the $R$\,=\,100 spectra between 4.2--4.8\,$\upmu$m, where the \jedi{} datapoints consistently sit at higher transit depths than the \tiberius data. All of our tests at the reduction level demonstrate that the data are robust to the reduction method used. However, as with the previous BOWIE-ALIGN targets, we test our models in the following sections on each reduction to assess the implications on the interpretation of this planet's atmosphere. 

\begin{figure}
    \centering
    \includegraphics[width=\linewidth]{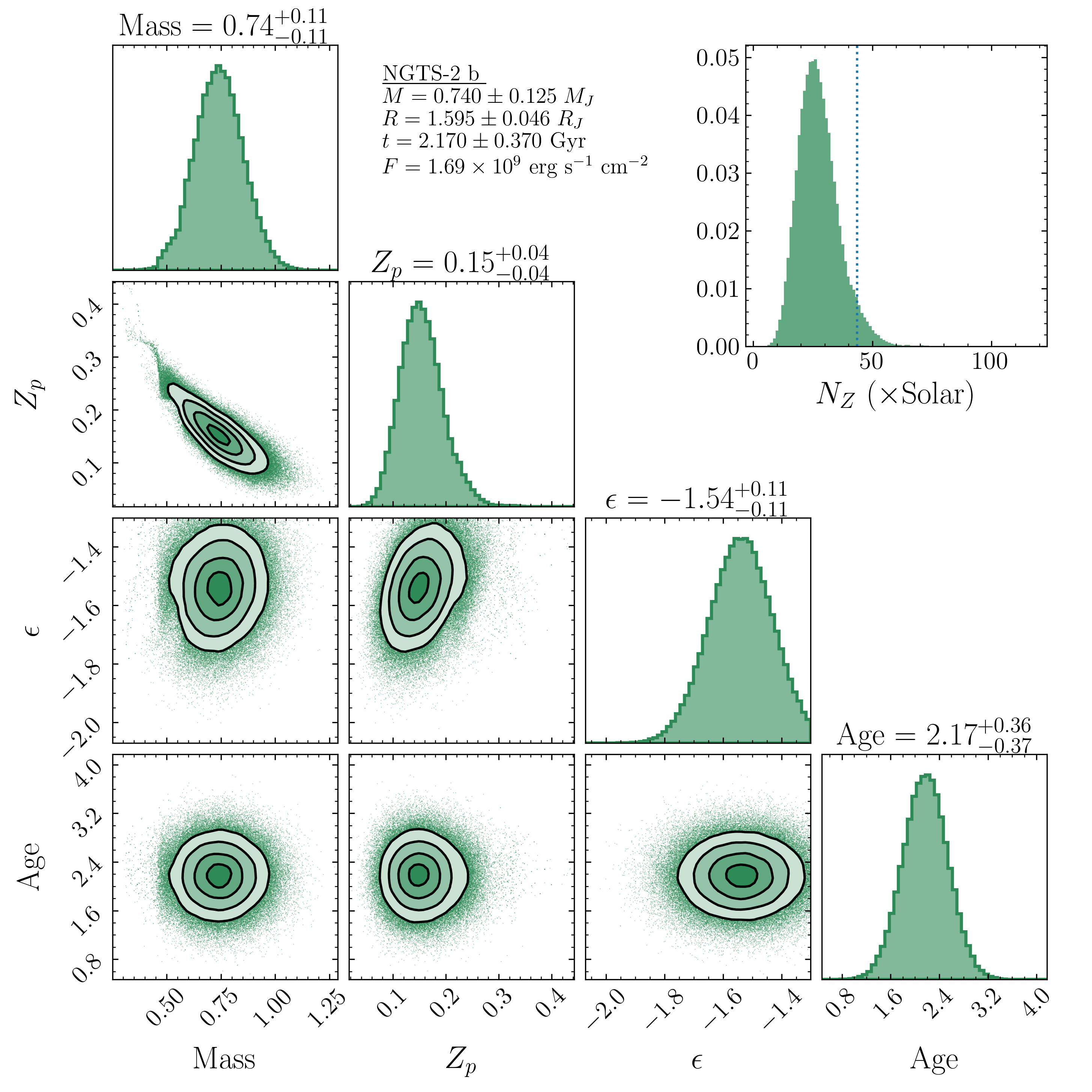}
    \caption{The posterior distribution of our interior structure modelling parameters: mass ($M_\mathrm{J}$), metallicity, log heating efficiency, and age.  The parameters used in the model are listed in the figure.  We infer a metallicity of $Z_\mathrm{p}=0.15\pm0.04$, below average for this mass but well within the natural dispersion seen in \citep{Thorngren2016}.  Converting this to a number ratio and assuming the limiting fully-mixed case, we obtain an upper-limit on the atmospheric metallicity of 43.5 $\times$Solar.}
    \label{fig:interior}
\end{figure}

\section{Interior Structure Models}\label{sec:interior}
To contextualize and constrain the atmospheric models, we first model the interior of the planet. In the first BOWIE-ALIGN paper, retrievals were drawn to high mean molecular weights, and interior structure models were implemented to provide more informed priors. We utilise the same approach here.  \emph{A priori}, we would expect a planet of this mass and equilibrium temperature to comprise roughly 20\% metals by mass \citep{Thorngren2016} and possess a radius of 1.33~$R_\mathrm{J}$ \citep[via][]{Thorngren2021}.  The planet is substantially larger than this ($R_\mathrm{p}\ \sim$ 1.6\,$R_\mathrm{J}$), suggesting that the planet either has a hotter or more metal poor interior than comparable planets.

Using the Bayesian retrieval method of \citet{Thorngren2019}, we fit interior structure models to NGTS-2b's observed mass, radius, and age, accounting for observational uncertainties and fixing incident stellar flux ($1.69\times10^9 \mathrm{erg}\,\mathrm{s}^{-1}\mathrm{cm}^{-2}$), calculated from $R$\,=\,400 reduction.  The resulting posterior is shown in Figure \ref{fig:interior}.  The additional parameter $\epsilon$ accounts for the anomalous hot Jupiter heating as the base-10 log of the fraction of incident flux injected into the interior of the planet, following \citep{Thorngren2018} -- it must be a parameter to account for the uncertainty in the flux-heating relationship.

We find that the planet is likely slightly more metal poor than comparable planets at $Z_\mathrm{p}\,=\,0.15\pm0.04$.  Note that these uncertainties account only for observational error, not possible modelling error.  An upper limit on the atmospheric metallicity may be derived from this as the limiting case of a fully mixed planet \citep{Thorngren2019}, shown in the top right of Figure \ref{fig:interior} as a number ratio ($\times$ solar).  This yields a 95\% upper limit on the atmospheric metallicity of 43.5\,$\times$ solar, where our solar value is defined as the number ratio for metals to hydrogen of $0.104\%$.

\section{Retrieval modelling}\label{sec:retrievals}

Taking the \jedi{} $R$\,=\,400 reduction as our primary spectrum for analysis, we conduct a retrieval exploration of the chemical inventory and pressure-temperature (P-T) structure of NGTS-2b. We use \texttt{v1.2} of the retrieval package \poseidon{} \citep{macdonaldmadhusudhan2017poseidon, macdonald2023poseidonjoss} to run a suite of model set-ups with both free and equilibrium chemistry and evaluate the most favoured frameworks based on the data. We compare these results to the \jedi{} $R$\,=\,100 spectra, and \tiberius{} $R$\,=\,400 and $R$\,=\,100 spectra to assess the robustness of our solutions between data reductions and resolutions. We obtain an additional set of retrievals using \prt{} \citep{molliere2019prt}, to assess the robustness of the inferred atmospheric properties to modelling assumptions. 

\subsection{\poseidon{} set up}

\renewcommand{\arraystretch}{1.2}
\begin{table}
    \begin{threeparttable}
    \caption{Fixed stellar and planetary parameters, alongside prior ranges of free parameters used in our \poseidon{} and \prt{} retrieval modelling. $\mathcal{U}$ indicates a uniform prior range and $\mathcal{N}$ indicates a Gaussian prior, where we specify the mean and standard deviation. Note that the parameter species abundances in \poseidon{} are defined as volume mixing ratios, whereas \prt{} specifies mass fractions.}
    \label{tab:retrieval_priors}
    \begin{tabular}{l c c c c c} 
    \hline
    Parameter & \poseidon{} & \prt{} \\
    \hline 
    \textit{(Fixed Parameters)} & \multicolumn{2}{c}{\citep{raynard2018ngts2b}} \\
    $R_\mathrm{*}$($R_\odot$) & \multicolumn{2}{c}{1.702} &\\
    T$_\mathrm{*}$ (K) & \multicolumn{2}{c}{6478} &\\
    $\log{g_\mathrm{*}}$ (cgs) & \multicolumn{2}{c}{4.197} & \\
    $[\mathrm{Fe/H}]_\mathrm{*}$ & \multicolumn{2}{c}{-0.06} &\\
    $M_\mathrm{P}$($M_\mathrm{J}$) & \multicolumn{2}{c}{0.74} & \\
    $R_\mathrm{P}$($R_\mathrm{J}$) & \multicolumn{2}{c}{1.595} & \\
    \hline
    \textit{(Priors)} \\
    $R_{\mathrm{P_{ref}}}$ ($R_\mathrm{P}$) & $\mathcal{U}$(0.85, 1.15) & $\mathcal{U}$(0.8, 2.2)\\
    $\log{g}$ & $\mathcal{N}$(2.858, 0.026)$^a$ & $\mathcal{N}$(2.858, 0.026)$^a$ \\
    $T$ (K) & $\mathcal{U}$(400, 2300) & $\mathcal{U}$(500, 3000)\\
    $\log{\mathrm{P_{cloud}}}$ (bar) & $\mathcal{U}$(-7, 2)  & $\mathcal{U}$(-6, 2)\\
    $\delta_\mathrm{rel}$ (ppm) & $\mathcal{U}$(-1000, 1000)& $\mathcal{U}$(-200, 200)\\
    \hline
    \textit{(Equilibrium Chemistry)} \\
    $[\mathrm{M/H}]$ & $\mathcal{U}$(-1, 1.7)$^b$ & $\mathcal{U}$(-2, 3)\\
     C/O & $\mathcal{U}(0.2, 1.2) $& $\mathcal{U}(0.1, 1.5) $\\
    \hline
    \textit{(Free Chemistry)} & \textit{VMR} & \textit{Mass Fraction} \\
    $\log{X}$ & $\mathcal{U}$(-12, -1) & $\mathcal{U}$(-12, -0.5) \\
    \hline
    \end{tabular}
    \begin{tablenotes}
    \small
    \item $^a$ $\log{g}$ and $\log{g_\mathrm{err}}$ are calculated from the $M_\mathrm{P}$ and $R_\mathrm{P}$ fixed parameter values.
    \item $^b$ A prior range of $\mathcal{U}$(-1, 3) is initially explored. Due to the presence of a high metallicity mode, the upper prior limit is restricted to 50$\times$ solar ([M/H] = 1.7).
    \end{tablenotes}
    \end{threeparttable}
\end{table}

We initiate a \ce{H}/\ce{He} dominated atmosphere with a fixed \ce{He}/\ce{H} volume ratio of 0.17, comprised of 100 pressure levels distributed uniformly in log space, spanning $10^2$ --$10^{-8}$ bar. The abundances of trace gases are assumed to be uniform in height, measured by their $\log_{10}{X}$ volume mixing ratios (VMRs). Our base inventory of chemical species consists of: \ce{H2O} \citep{polyansky2018h2o}, \ce{CO} \citep{li2015co}, \ce{CO2} \citep{yurchenko2020co2}, \ce{CH4} \citep{yurchenko2024ch4}, \ce{NH3} \citep{coles2019nh3} \ce{HCN} \citep{barber2014hcn}, \ce{H2S} \citep{azzam2016h2s}, \ce{SO2} \citep{underwood2016so2}, and \ce{SO} \citep{brady2024so}. To compute the radiative transfer in \poseidon{}, we define a model wavelength grid spanning 2.6--5.2\,$\upmu$m at a resolution of $R\,=\,30\,000$, from which models are binned to the resolution of the observations and convolved with the NIRSpec/G395H transmission function. The parameter space is evaluated using the nested sampling package \texttt{PyMultiNest} \citep{buchner2014pymultinest} at a resolution of 1000 live points to obtain marginalized posterior distributions and evaluate model goodness-of-fit.

The stellar and planetary parameters are fixed to the values presented in {Table \ref{tab:retrieval_priors}} except for $\log{g}$, which is allowed to vary according to a Gaussian prior with a standard deviation equivalent to the measured error. The reference pressure is set at 10 bar, from which the radius at the reference pressure ($R_\mathrm{P,\,ref}$) is allowed to vary as a free parameter. Prior distributions for each model parameter are also specified in Table \ref{tab:retrieval_priors}. 

To evaluate model complexity, we test a series of cloud parametrizations and P-T profiles, given the evidence indicating pressure-temperature profile complexity can counteract biases in atmospheric retrievals \citep[e.g.,][]{lueber2024informationcontent, schleich2024ptprofile}. 
We test (i) an isothermal, cloud-free atmosphere, (ii) an isothermal atmosphere with a homogeneous grey cloud deck, and (iii) a 4-parameter Guillot P-T profile \citep{Guillot2010pt} with a homogeneous grey cloud deck. For each of the parameterisations outlined, we test models with and without a detector offset where NRS2 is fixed and NRS1 is allowed to vary with a uniform prior. 
From our retrievals, we find the inclusion of \ce{SO} opacity critically changes the retrieved atmosphere (see section \ref{sec:poseidon_results}), so these tests are performed on models both including and excluding \ce{SO} as a trace species.

\begin{figure*}
    \centering
    \includegraphics[width=\linewidth]{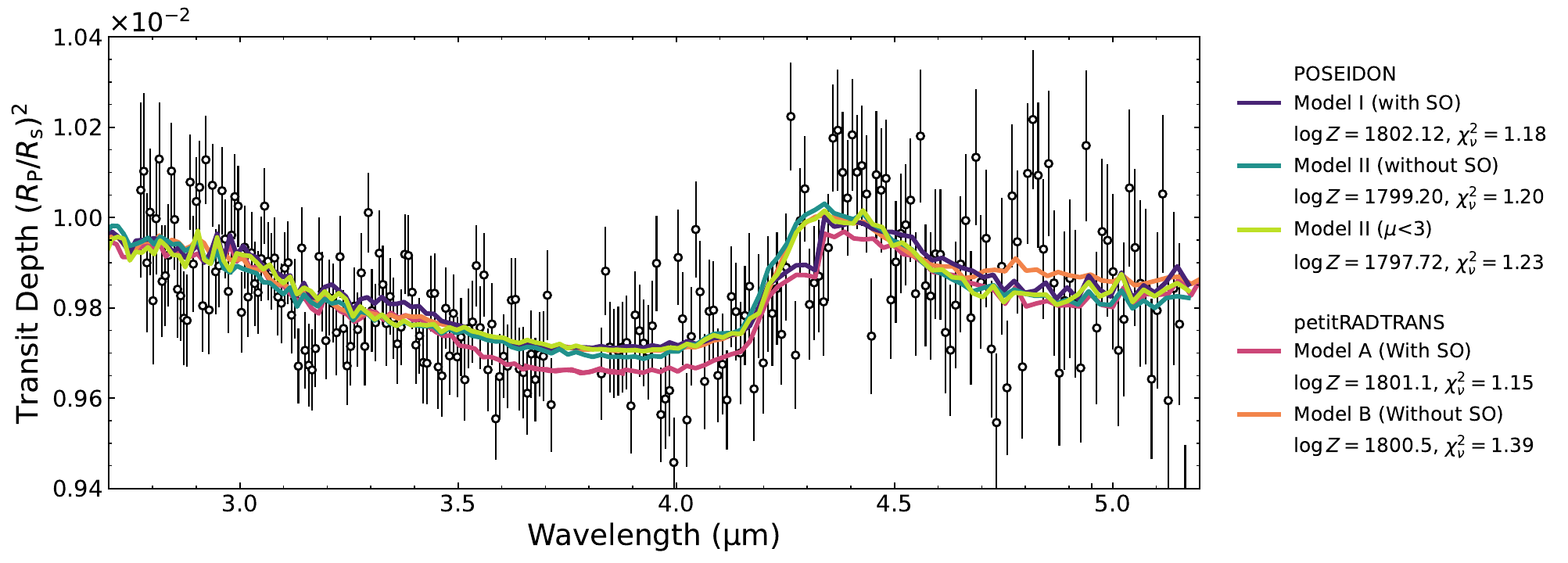}
    \includegraphics[width=\linewidth]{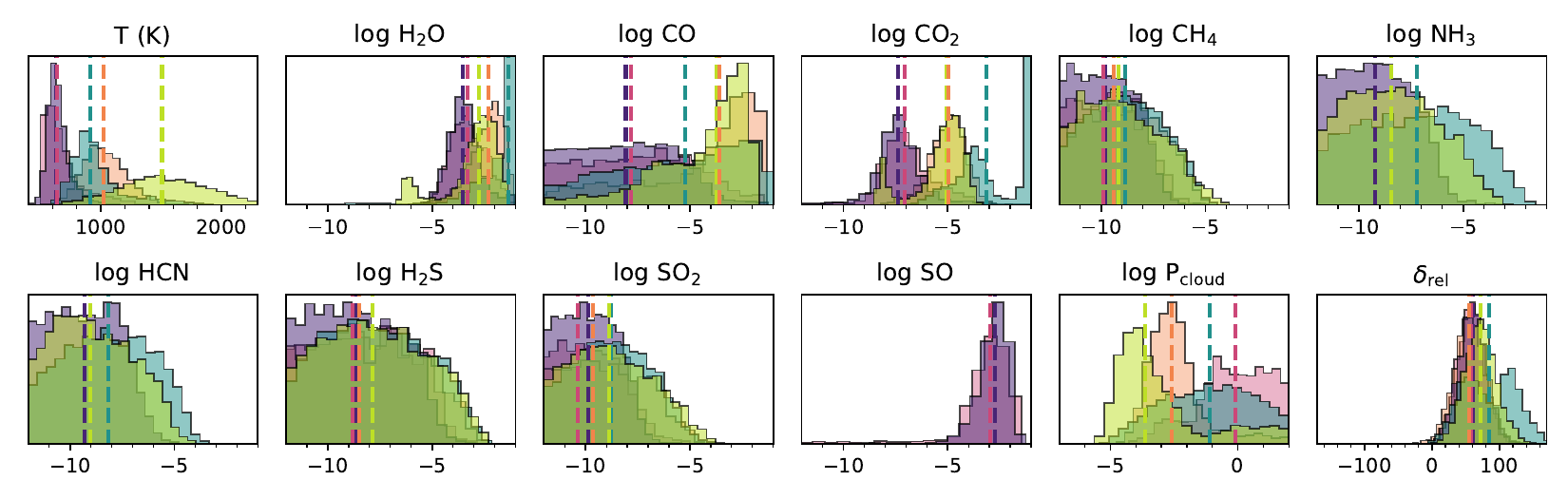}
    \caption{Free chemistry retrievals on the \jedi{} $R$\,=\,400 spectrum for the three models tested with \poseidon{} (Models I--III) and two models tested with \prt{} (Models A and B). Model I and Model A including SO opacity (purple and pink). Model II and Model B excluding SO opacity (blue and orange). Model II excluding SO opacity and limiting the mean molecular weight to $\upmu\,<\,3$ (light green). While Model I statistically favoured ($\Delta \log Z$\,=\,5) over Model II ($\upmu\,<\,3$), we discuss the astrophysical likelihood of the retrieved abundance of SO in Section\,\ref{sec:SOchem}, and the role of the mean molecular weight in Section\,\ref{sec:mmw}. We note that due to \poseidon{} and \prt{} retrieving offsets for NRS1 and NRS2 respectively, the \prt{} offset has been inverted to directly compare to the \poseidon{} retrieved value.}
    \label{fig:poseidon_spec_hist}
\end{figure*}

Of these six model variations, for atmospheres with and without \ce{SO}, the greatest difference in the log evidence between models is 2.25 and 1.06 respectively. For models with only one difference in parameterisation (e.g. a different P-T profile), the difference generally falls close to, but below one, such that no model is statistically favoured over any other \citep{trotta2008stats}. From these permutations, we default to the highest evidence model as our base model for the atmosphere with \ce{SO} \textbf{(Model I)}; this is an isothermal cloud-free atmosphere with offsets. For the atmosphere without \ce{SO} \textbf{(Model II)}, we select an isothermal atmosphere with grey cloud opacity and an offset. Despite the cloud-free model providing a marginally higher evidence of fit, when including a cloud deck, the cloud pressure posteriors remain unconstrained. Therefore, we cannot state that this is a cloud-free atmosphere based on these data alone and consequently, we include a grey cloud deck to show the degeneracy and limits of cloud detection with this wavelength coverage. We note that the retrieved Guillot profiles retrieve consistent temperatures with isothermal profiles within the regions probed by transmission spectroscopy and therefore use the isothermal profile to reduce the number of fit parameters. The retrieved transmission spectra and posterior distributions of Models I and II are displayed in Figure \ref{fig:poseidon_spec_hist}. 

To apply equilibrium retrievals, we modify the free retrieval setup by reducing the minimum pressure of the atmosphere grid to $10^{-7}$ bar (the minimum pressure of the \texttt{Fastchem} \citep{stock2018fastchem, stock2022fastchem} equilibrium chemistry grid implemented within \poseidon{}). We include the trace species \ce{H2O}, \ce{CO}, \ce{CO2}, \ce{CH4}, \ce{NH3}, \ce{HCN}, \ce{H2S} and \ce{SO2}. \ce{SO} is not included as it is not present in the \texttt{Fastchem} grid. Metallicity and C/O are allowed to vary as free parameters (see Table \ref{tab:retrieval_priors} for prior ranges) where the C/H ratio is derived from the O/H and C/O ratios. We find no evidence to support the inclusion of a P-T profile more complex than isothermal, leading to a seven parameter retrieval with a grey cloud deck and offset between NRS1 and NRS2.

\subsection{\prt{} set up}

Following the analysis method outlined for the BOWIE-ALIGN survey \citep{Kirk2024}, we perform an additional set of independent retrievals using the \prt \citep[v3.1.3,][]{molliere2019prt, Nasedkin2024, blain2024} package, with a setup similar to that used for other BOWIE-ALIGN targets including WASP-15b \citep{kirk2025_w15} and TrES-4b \citep{meech2025_t4}. \prt uses Bayesian nested sampling from \texttt{pyMultiNest} \citep{buchner2014pymultinest} to explore the parameter space of model transmission spectra. We use free, equilibrium, and hybrid chemistry retrieval set-ups on both the \jedi{} and \tiberius{} data reductions at $R$\,=\,100 and $R$\,=\,400. 

For the radiative transfer calculations of model transmission spectra, we use $R$\,=\,1,000 correlated-$k$ line opacities from CO \citep{Rothman2010_hitemp}, \ch{H2S} \citep{azzam2016h2s}, \ch{SO2} \citep{underwood2016so2}, \ch{CH4} \citep{yurchenko2024ch4}, \ch{H2O} \citep{polyansky2018h2o}, \ch{NH3} \citep{coles2019nh3}, \ch{CO2} \citep{yurchenko2020co2}, \ch{HCN} \citep{barber2014hcn}, and SO \citep{brady2024so}, plus Rayleigh scattering from \ch{H2} and He, and collisionally-induced absorption from \ch{H2-H2} and \ch{H2-He} \citep{borysow1988collison,borysow2001high,borysow2002collision}. We perform the calculation over 100 equally log-spaced pressure layers from $10^{-6}$ to $10^2$  bar, with the stellar radius fixed to 1.702 $R_{\odot}$ from \citet{raynard2018ngts2b}, and the reference radius set at a pressure of 1 mbar.

We use an isothermal temperature-pressure profile with a wide uniform prior on isotherm temperature of 500--3000\,K, a uniform prior on reference radius of 0.8--2.2 $R_\mathrm{J}$, and a Gaussian prior on the gravity calculated from the mass and radius presented in \citet{raynard2018ngts2b}. We use a grey cloud deck with a log-uniform prior on cloud-top pressure from $10^{-6}$ to $10^2$ bar (any altitude in the modelled atmosphere) to parametrize the impact of aerosols. We permit for an offset between the detectors in our retrievals, with a uniform prior of up to 200\,ppm in either direction. Our full retrieval priors are outlined in Table \ref{tab:retrieval_priors}.

\begin{figure}
    \centering
    \includegraphics[width=\linewidth]{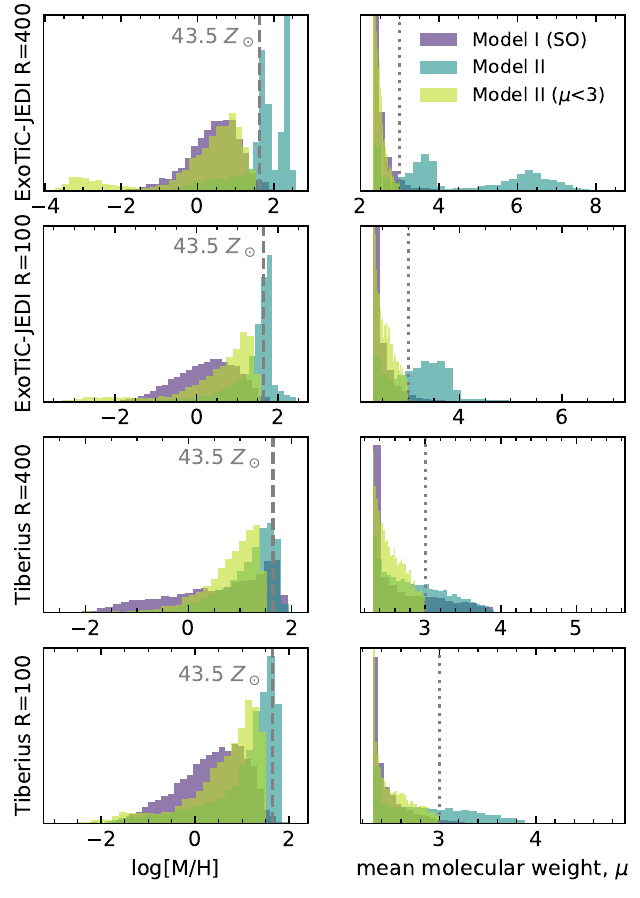}
    \caption{Probability density histograms comparing each of our three free chemistry \poseidon{} models by retrieved metallicity and mean molecular weight across our two resolutions and two reductions. Across both resolutions and reductions we find similar results in each of our models, with the exception of \jedi{} $R$\,=\,400 which shows a multi-modal solution for Model II in both metallicity and $\upmu$.}
    \label{fig:mmw_hist}
\end{figure}

In our free chemistry retrievals, the abundances of \ch{CH4}, \ch{H2O}, \ch{CO}, \ch{CO2}, \ch{H2S}, \ch{SO2}, and SO, are free parameters (\textbf{Model A}), with a wide log-uniform prior on mass fraction from $10^{-12}$ to $10^{-0.5}$, and the remaining atmosphere assumed to be a solar mixture of  \ch{H2} and He. 
We also include retrievals without SO on the $R$\,=\,400 spectra (\textbf{Model B}). In our equilibrium chemistry retrievals, the atmospheric abundances of \ch{H2}, He, \ch{CH4}, \ch{H2O}, \ch{CO2}, CO, \ch{NH3}, \ch{H2S}, and HCN are set by equilibrium chemistry calculated given the temperature, C/O ratio, and metallicity in each pressure layer. 
The C/O ratio and metallicity are free parameters with wide, log-uniform priors of 0.1--1.5, and $10^{-2}$ to $10^{3}\times$ solar respectively. The equilibrium parameterisation used by \prt fixes the C/H ratio based on the metallicity, relative to solar \citep{Asplund2009}, with the O/H ratio set by the C/H ratio divided by the C/O ratio. Our hybrid chemistry set-up combines these two approaches, using equilibrium chemistry for all non-sulphur-bearing species, while the abundances of \ch{H2S}, \ch{SO2}, and SO are free parameters. 
This permits a broadly equilibrium chemistry interpretation, while also allowing the sulphur species to freely vary, potentially capturing the effects of photochemical creation of \ch{SO2} and SO and depletion of \ch{H2S}, or primordial enhancement or depletion of the O/S ratio.

\begin{figure*}%
    \centering
    \subfloat[\poseidon{}]{{\includegraphics[width=0.48\textwidth]{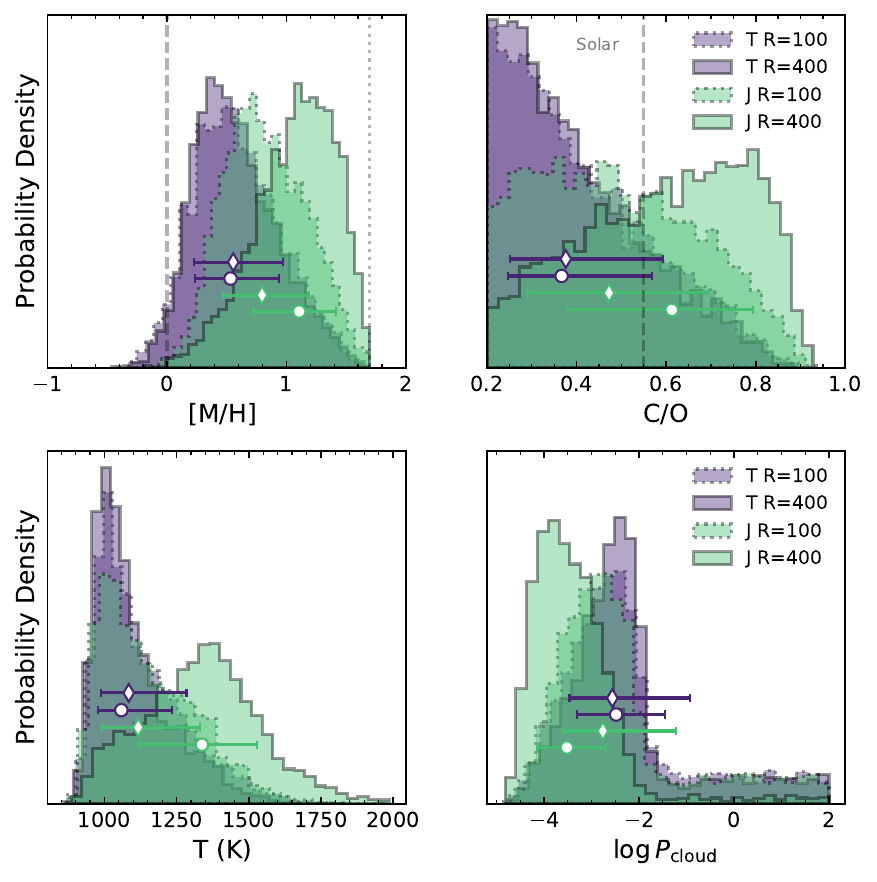} }}%
    \subfloat[\prt{}]{{\includegraphics[width=0.48\textwidth]{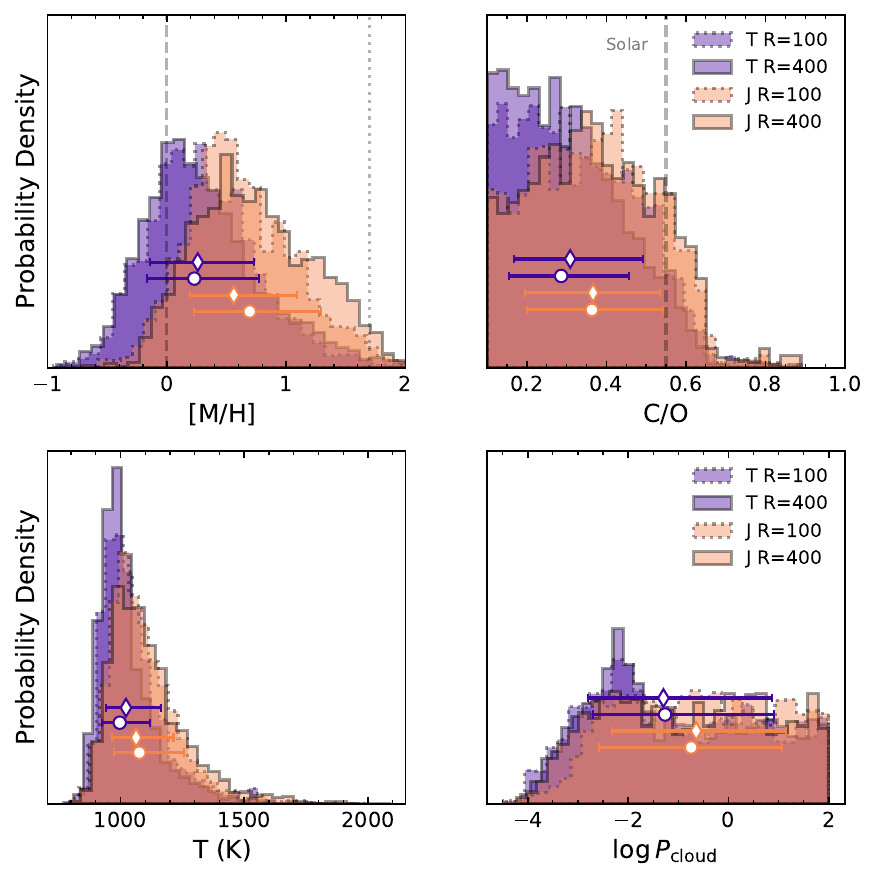} }}%
    \caption{Probability density histograms for our Equilibrium \poseidon{} retrievals \textbf{(left)} and \prt{} retrievals \textbf{(right)} on each reduction and resolution of the data. For our \poseidon{} retrievals, We find that all models point to a solar to super-solar metallicity and C/O abundances skewed towards low values (with the expection of the \jedi{} $R=400$ reduction), and a relatively high cloud deck. For our \prt{} retrievals, the metallicity spans solar to supersolar values with subsolar C/O ratios favoured and no constraints on the pressure level of the cloud deck.}%
    \label{fig:eqchem_hist}%
\end{figure*}

\renewcommand{\arraystretch}{1.2}
\setlength{\tabcolsep}{4pt}
\begin{table}
    \caption{Detection significances of H$_2$O, CO, CO$_2$, and SO on the \jedi{} R=400 reduction. We report the Bayes factor as $\log_{10}\mathcal{B}_{1,2}$, where a value between 1-2 indicates strong evidence, and values $>$ 2 indicate decisive evidence in favour models including the given species following the categorisation in \citet{kass1995bayes}. Corresponding sigma significances are computed using equation 9 of \citet{thorngren2025bayes}. No other species tested show detection significances greater than 1$\sigma$. Given the lack of physical plausibility of Models I and II, and the strict prior imposed on Model II ($\mu < 3$), we caution the reader against quoting these significances without appropriate context. 
    }
    \label{tab:detection_sigs}
    \begin{tabular}{l c c c c c c c c c c} 
    \hline
     Retrieval & \ce{H2O} & & \ce{CO} &  & \ce{CO2} & & \ce{SO} & &\\
     \textit{\poseidon{} $R$=400} &  $\mathcal{B}_{1,2}$ & $\sigma$ & $\mathcal{B}_{1,2}$ & $\sigma$ & $\mathcal{B}_{1,2}$ & $\sigma$ & $\mathcal{B}_{1,2}$ & $\sigma$ & \\
     \hline
     \textit{\textbf{Model I} (SO)} & 1.61 & 2.3 & 0.02 & 0.7 & 1.19 & 1.9 & 1.32 & 2.0 \\
     \textbf{\textit{Model II }} & 1.15 & 2.8 & 0.30 & 1.8 & 5.59 & 5.4 & - & - & \\
     \textit{\textbf{Model II} ($\mu < 3$) } & 4.16 & 4.0 & 0.53 & 1.2 & 5.80 & 4.8 & - & -\\
     \hline
     \\
    \end{tabular}

\end{table}

\subsection{Free chemistry results}\label{sec:poseidon_results}

\textbf{Model I:} For our \poseidon{} models including \ce{SO} opacity, we find weak evidence of absorption from \ce{H2O}, \ce{CO2} and \ce{SO} (Figure \ref{fig:poseidon_spec_hist}). 
We retrieve an extremely sulphur-enriched atmosphere, with a median retrieved abundance from the $R$\,=\,400 retrieval of $\log{X_\mathrm{SO}} = -2.73^{+0.57}_{-0.77}$, despite finding no evidence for \ce{SO2} opacity (see Discussion \ref{sec:SOchem}). 
Comparatively, the median abundances of $\log{X_\mathrm{H_2O}} = -3.51^{+0.73}_{-0.67}$ and $\log{X_\mathrm{CO_2}} = -7.38^{+0.94}_{-0.81}$, align more closely with solar equilibrium expectations. 
The retrieved temperature of $T=633^{+93}_{-65}$ K is significantly lower than the planetary equilibrium temperature of 1468 K (we discuss this further in Section \ref{sec:SOchem}), marking a cold, sulphur-rich atmosphere with a metallicity of $3.16^{+8.86}_{-2.59}\times$ solar. Comparing the Bayesian evidence between Model I and II (with and without \ce{SO}), we find the presence of \ce{SO} is statistically favoured at a significance of $2.0 \sigma$, with a $\chi^2_\nu$ of 1.18 (Tables \ref{tab:detection_sigs}, \ref{tab:retrieval_results_main}). This does not indicate a robust detection, and it is important to interpret the results of free chemistry retrievals, which lack underlying physical motivation, with caution. To this effect, we interrogate why such an atmosphere can provide a good fit to the data, and the plausibility of a sulphur enriched, low temperature atmosphere in section \ref{sec:SOchem}.

Across the remaining reductions (\jedi{} $R$\,=\,100, \tiberius{} $R$\,=\,400 and $R$\,=\,100), the retrieved parameters are consistent to within 1$\sigma$ (Tables \ref{tab:retrieval_results_main} and \ref{tab:retrieval_results_free_abundances}), although the \tiberius{} $R$\,=\,400 spectrum finds a high abundance \ce{H2O} and \ce{CO2} mode. Similarly, the \prt{} retrievals (e.g., Model A, Figure \ref{fig:poseidon_spec_hist}) on all four reductions converge on the same high \ce{SO} abundance, low temperature solution, with no evidence for cloud opacity, with a 2$\sigma$ upper limit of 10 mbar.

\textbf{Model II:} Removing \ce{SO} from the \poseidon{} retrieval set-up leads to a high mean molecular weight atmosphere, driven by high abundance modes of $\log{X_\mathrm{H_2O}} = -1.33^{+0.25}_{-1.39}$ and $\log{X_\mathrm{CO_2}} = -3.15^{+2.09}_{-1.33}$ which converge on the upper prior bounds. This leads to a median metallicity of $65^{+219}_{-49}\times$ solar with 72$\%$ of the posterior distribution falling above the 95$\%$ upper metallicity limit of 43.5$\times$ solar from our interior structure models (see Figure \ref{fig:mmw_hist}). A temperature of $916^{+236}_{-131}$K is retrieved with no constraints found for the cloud pressure, and the retrieved offset shows a bimodality with peaks around 60 and 110\,ppm, where the high mean molecular weight is more pronounced within the 110\,ppm mode (see Figure \ref{fig:poseidon_spec_hist_m2}). This peak occurs where \ce{CO2} replaces \ce{H2O} as the species responsible for reducing the atmospheric scale height through the mean molecular weight. Whilst still hosting a high mean molecular weight solution due to a high \ce{H2O} abundance, the 60\,ppm mode finds a \ce{CO2} log VMR of $\sim -4$, producing an atmosphere similar to one retrieved with no offset. 

Like the \jedi{} $R$\,=\,400 reduction, the \jedi{} $R$\,=\,100, and \tiberius{} $R$\,=\,100 and $R$\,=\,400 reductions all converge towards high mean molecular weight modes through the \ce{H2O} abundance, but lack the high \ce{CO2} mode that drives the larger offset between NRS1 and NRS2. The \tiberius{} $R$\,=\,400 reduction shows the strongest preference for a cloud deck with a median cloud pressure level of $\log{P_\mathrm{cloud}} 
\,=\,-2.5^{+1.9}_{-0.7}$ bar. Remaining differences between retrieved parameters are minimal. This solution of a high \ce{H2O} abundance is also supported within the \prt{} retrievals, where the retrieved \ce{H2O} mass fraction is equivalent to a mixing ratio of $\sim4\%$ and a total mean molecular weight of 4.4. The \ce{CO2} abundance is consistent with the \poseidon{} 60\,ppm offset mode, with an equivalent mixing ratio of $\log{X_\mathrm{CO2}} = -4.90^{+0.71}_{-0.87}$, temperature of $914^{+174}_{-145}$~K, and cloud deck at $\log{P_\mathrm{cloud}} = -2.43^{+1.11}_{-0.72}$.

Such a preference for high abundance modes of trace gases develops by reducing the scale height of the atmosphere to fit spectral features via the mean molecular weight. 
Given the expected upper metallicity limit of the atmosphere from interior structure modelling, we restrict the allowed mean molecular weight of models within our retrieval to values of $\mu < 3$ to investigate whether alternative solutions emerge.

\textbf{Model II ($\mu<3$) - restricting the mean molecular weight:} With an upper mean molecular weight limit of 3 placed on the retrieval prior, we report abundances of $\log{X_\mathrm{H_2O}} = -2.75^{+0.74}_{-1.63}$, $\log{X_\mathrm{CO}} = -3.69^{+1.22}_{-3.07}$ and $\log{X_\mathrm{CO_2}} = -5.07^{+0.74}_{-1.73}$, where the inclusion of \ce{H2O} and \ce{CO2} are supported at significances of 4.0$\sigma$ and 4.8$\sigma$, respectively. 
The retrieved isothermal temperature is poorly constrained, yet consistent with the equilibrium temperature at $1507^{+379}_{-349}$K and a cloud deck emerges at a median log-pressure of $-3.63^{+2.72}_{-0.92}$. However, under this prior restriction, a bimodality exists in the retrieved atmosphere, driven by the presence of or lack of cloud in the observed spectrum. The median atmosphere is characterised by the cloudy solution, however, a cloud-free, low metallicity atmosphere is also allowed.

In contrast with the additional reductions, only the \jedi{} $R$\,=\,400 reduction finds this bimodality between clouds and metallicity. The remaining reductions only support a high metallicity atmosphere and span a range of median temperatures, between 850--1000\,K. Only the \jedi{} reductions favour a high \ce{CO} abundance and all reductions support the presence of cloud opacity, with median cloud pressure levels spanning from $10^{-3.6}$--$10^{-1.8}$ bar. Comparing the metallicity distribution of all reductions (Figure \ref{fig:mmw_hist}), only the \jedi{} $R$\,=\,400 reduction appears to decay at high metallicities, although this may be an artifact of enhanced \ce{CO} abundance replacing \ce{H2O} at high mean molecular weights. All other reductions can easily be seen as truncated distributions of Model II.

Given that our abundances remained skewed towards the upper prior bound, our retrievals are unable to find an alternative, high likelihood solution from restricting the parameter space by mean molecular weight. Median retrieved parameters with 1$\sigma$ errors are reported for Model I and Model II (for both the unrestricted and restricted mean molecular weight retrievals) in Tables \ref{tab:retrieval_results_free_abundances} and \ref{tab:retrieval_results_main}, along with marginalised posterior distributions in Appendix \ref{sec:cornerplots}. Detection significances, calculated using equation 9 of \citet{thorngren2025bayes}, are given in Table \ref{tab:detection_sigs}.  We caution the reader against quoting these values and detection significances outside of the context of the priors placed on these parameters, and plausibility of the retrieved solutions.

\begin{figure*}
    \centering
    \subfloat[\poseidon{}]{{\includegraphics[width=0.48\textwidth]{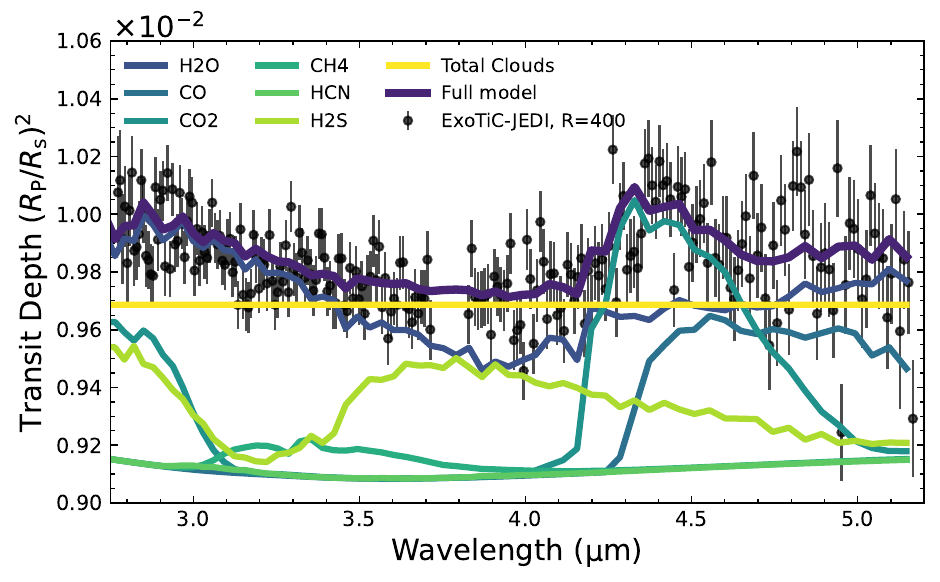} }}%
    \subfloat[\prt]{{\includegraphics[width=0.48\textwidth]{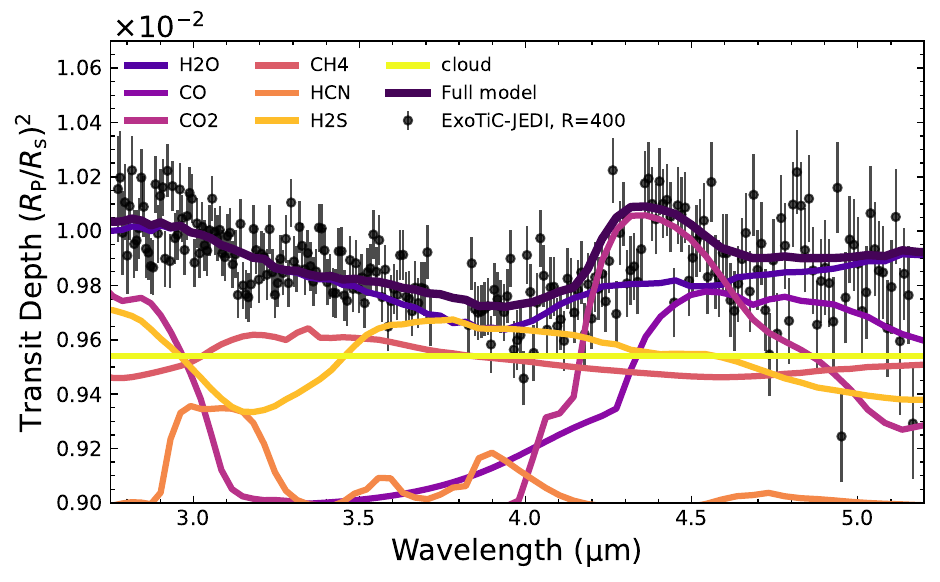} }}%
    \caption{Spectral contributions of the median retrieved atmospheres from the \poseidon{} (\textbf{left}) and \prt (\textbf{right}) retrievals on the \jedi{} $R$=400 reduction. Although weak constraints are found from our equilibrium chemistry retrievals, the contributions show how the muting of the \ce{H2O} and \ce{CO2} spectral features is being driven by cloud opacity in \poseidon{} and \ce{H2S} opacity in \prt{}.}
    \label{fig:pos_prt_equilibrium_contributions}
\end{figure*}

\subsection{Equilibrium chemistry results}

For the \poseidon{} equilibrium chemistry retrievals, we initially adopt wide, uninformative priors on the metallicity of $\mathcal{U}(-1,3)$. However, this leads to a bimodal solution with a dominant, high metallicity mode at $\sim 200\times$ solar for the \jedi{} $R$\,=\,400 spectrum, where a wide range of cloud pressure levels are possible. However, lower cloud pressures ($<10^{-2}$ bar) are necessary to fit the spectrum at lower metallicities. This high metallicity mode is also present, albeit weakly, in the Tiberius $R$\,=\,400 and $R$\,=\,100 spectra. From our interior structure modelling, we expect the 95$\%$ upper limit on atmospheric metallicity to be $43.5\times$ solar. Therefore, we adjust our prior upper limit to $50\times$ solar $\mathcal{U}(-1, 1.7)$ to account for this.

When applying the metallicity limited prior, our \poseidon{}{} retrievals support a range of metallicities and C/O ratios (Figure \ref{fig:eqchem_hist}), with the median retrieved [M/H] spanning from 3.4$\times$--12.9$\times$ solar, and C/O spanning 0.38--0.61 between reductions. Despite this, all retrievals overlap at the 1$\sigma$ level due to weak constraints on the atmospheric parameters. Overall, the retrievals support the presence of a super-solar metallicity atmosphere, with the \tiberius{} reductions finding lower metallicities than \jedi{}. With the exception of the \jedi{} $R$\,=\,400 reduction, the remaining reductions skew towards lower C/O ratios, although solar C/O cannot be ruled out at the 1$\sigma$ level. 
The \jedi{} $R$\,=\,400 reduction is also the highest metallicity case, marked by a higher retrieved temperature than the remaining reductions at $1340^{+191}_{-219}$K (compared to the \tiberius{} $R$\,=\,400 reduction at $1058^{+175}_{-80}$K).
All reductions support the need for a high altitude cloud deck, with the 1$\sigma$ limits across retrievals spanning pressures of $10^{-4.15}$--$10^{-1.40}$ bar, to reduce the size of spectral features. 

Our \prt{} equilibrium chemistry results are consistent between reductions, demonstrating a temperature of $\sim1000$\,K, and the same $56\pm20$ ppm offset found in the free retrieval. Unlike the \poseidon{} results, poor constraints are retrieved on the cloud deck altitude. The C/O ratio is consistently sub-solar, at $0.36^{+0.18}_{-0.16}$ in the \jedi{} $R$\,=\,400 reduction and $0.29^{+0.17}_{-0.13}$ in the \tiberius{} $R$\,=\,400 reduction, while the metallicity varies somewhat, with the lowest value of [M/H] = $0.23^{+0.55}_{-0.39}$ in the \tiberius{} $R$= 400 reduction and the highest value of [M/H] = $0.69^{+0.59}_{-0.46}$ in the \jedi{} $R$\,=\,400 reduction. This amounts to a difference of the \jedi{} reduction possessing about twice the median oxygen and triple the median carbon content as that suggested by the \tiberius{} reductions. The hybrid retrieval results closely follow their equilibrium counterparts, with no evidence favouring the inclusion of any of the three considered sulphur species of \ch{H2S}, SO, or \ch{SO2}. 

Figure \ref{fig:pos_prt_equilibrium_contributions} breaks down the opacity contribution of gaseous species and clouds in the median retrieved \poseidon{} and \prt{} spectra for the \jedi{} $R$\,=\,400 observation. This highlights how the cloud deck is responsible for significant muting between the \ce{H2O} and \ce{CO2} features in the \poseidon{} spectrum whereas \ce{H2S} provides opacity that creates this same muting effect in the \prt{} spectrum. 

\begin{figure}
    \centering
    \includegraphics[width=\linewidth]{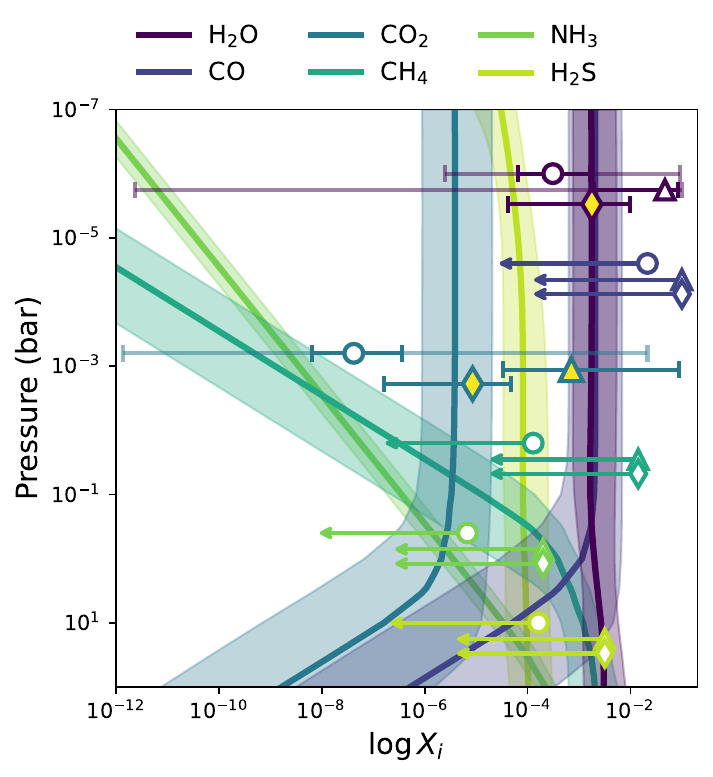}
    \caption{Chemical abundance plot showing the dominant retrieved species in NGTS-2b's atmosphere. Abundances from our equilibrium retrievals are calculated under equilibrium chemistry assumption using \texttt{Fastchem}, where we plot the median and 1$\sigma$ volume mixing ratios of our retrieved species across the \poseidon{} retrievals on all reductions. The $3\sigma$ upper limits for \ce{CO}, \ce{CH4}, \ce{NH3} and \ce{H2S} for all three \poseidon{} models, retrieved on the \jedi{} $R$\,=\,400 reduction, are overlaid. For species where at least one of the models has a detection significance $>3\sigma$ (Table \ref{tab:detection_sigs}), the median mixing ratio and 1$\sigma$ errors are plotted. For Models where the species has a detection significance $> 3\sigma$, the median marker is filled yellow, for models below the detection significance threshold, the full $3 \sigma$ bounds are also overlaid. Each of our free chemistry models are denoted as follows: circles = Model I, triangles = Model II, diamonds = Model II ($\upmu < 3$).
    }
    \label{fig:spaghetti}
\end{figure}

Finally, in Figure \ref{fig:spaghetti}, we plot the median retrieved equilibrium mixing ratios and $1\sigma$ confidence intervals from the combined \poseidon{} posteriors of all four reductions. Overlaid are the $3\sigma$ upper limits on the mixing ratios of chemical species without detections across all three \poseidon{} free chemistry models (Model I, Model II and Model II ($\mu<3$). For \ce{H2O} and \ce{CO2}, both of which have detection significances $>3\sigma$ in at least one of the models (Table \ref{tab:detection_sigs}), the median mixing ratio and $1\sigma$ errors are shown. Additionally, the $3\sigma$ bounds are displayed for the \ce{H2O} and \ce{CO2} abundances without $>3\sigma$ significances. For Model I (containing \ce{SO}, circles), although the \ce{H2O} mixing ratio is in agreement with equilibrium chemistry within its $1\sigma$ errors, the $1\sigma$ limits of \ce{CO2} are depleted relative to the retrieved equilibrium expectations. In contrast, \ce{CO2} is enhanced relative to equilibrium in Model II (triangles). The median \ce{H2O} abundance is also enhanced despite the $1\sigma$ tail of the distribution falling within equilibrium expectations. The $3\sigma$ bounds of the \ce{H2O} abundance show an unconstrained tail to the distribution, spanning the entire prior space. When imposing a mean molecular weight limit (Model II ($\mu < 3$)), \ce{H2O} and \ce{CO2} both fall within the retrieved equilibrium abundances although the 1$\sigma$ constraints span abundances beyond the retrieved equilibrium limits.

\begin{figure}
    \centering
    \includegraphics[width=\linewidth]{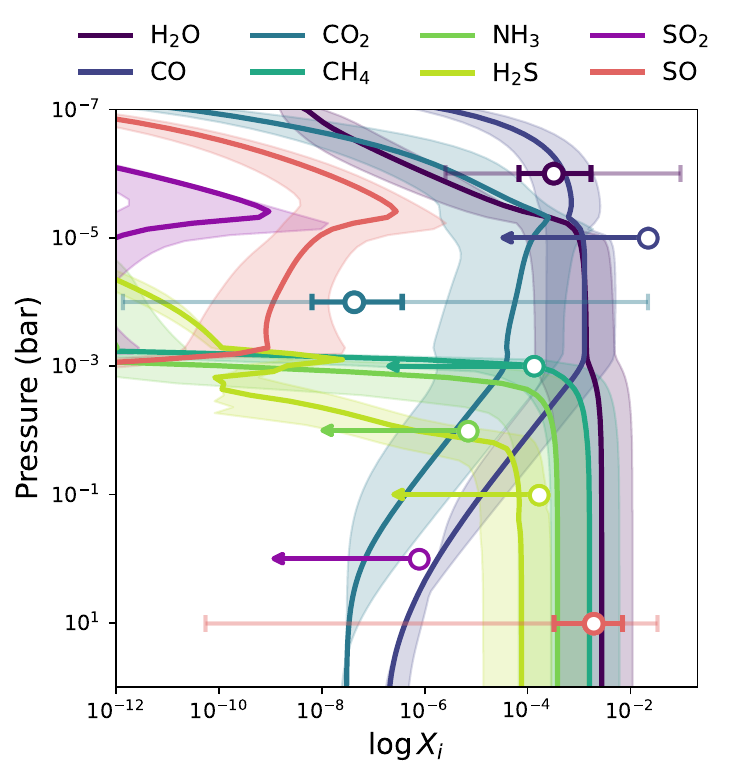}
    \caption{Same as Figure 9 but for steady-state abundances simulated using the photochemical model VULCAN. Abundances are calculated based on the retrieved isothermal temperature profile of 640 K from Model I. We use the median 3.16x solar metallicity for the corresponding median volume mixing ratios and 1$\sigma$ range of metallicity ($3.16^{+8.86}_{-2.59}\times$ solar) for the shaded region. Overlaid are our retrieved VMRs for our \jedi{} $R$=400, \poseidon{} Model I retrieval.}
    \label{fig:spaghetti_vulcan}
\end{figure}

\section{Discussion}\label{sec:discussion}

\subsection{Sulphur abundance expectations}\label{sec:SOchem}

Our \ce{SO} abundance from the \poseidon{} $R$\,=\,400 Model I retrieval is found to be $\log{X_\mathrm{SO}} = -2.73^{+0.57}_{-0.77}$, a result that is consistent across all reductions as well as in our pRT Model A run (see Figure\,\ref{fig:poseidon_spec_hist}). Given that it is unlikely for \ce{SO} abundances to be so high (around two orders of magnitude greater than solar abundances, assuming all sulphur is present in \ce{SO}), or for the retrieved atmospheric temperature to drop as low as $\sim$630\,K, we explore why we do not expect to see significant \ce{SO} opacity and why this model provides the best free-chemistry fit to the data.

Whilst the corresponding median metallicity of $3.16^{+8.86}_{-2.59}\times$ solar falls well within our interior structure expectations, the $[\mathrm{S/H}]$ is retrieved at $1.91^{+0.58}_{-0.77}$, suggesting an atmosphere that is highly sulphur enhanced, without significant enrichment of other species. 
\ce{SO} is photochemically produced through the oxidation of atomic sulphur by \ce{OH} radicals and is an intermediary species in the formation pathway of \ce{SO2} \citep{tsai2023, zahnle2009}. As such, \ce{SO} and \ce{SO2} abundances are linked. At 1000 K, abundances of the two species are comparable, yet as temperature increases, the relative abundance of \ce{SO2} to \ce{SO} decreases \citep{tsai2021, hobbs2021, polman2022}. Although quantitatively estimating this relative abundance would require photochemical modelling, with the equilibrium temperature range of 1000-1800 K, \ce{SO2} abundance is expected to follow well within two orders of magnitude of \ce{SO} \citep{tsai2023, hobbs2021}. Given the retrieved VMR of \ce{SO}, \ce{SO2} should therefore be present at observable abundances. The non-detection of \ce{SO2}, with a 2$\sigma$ upper limit of -7.49 dex, provides additional evidence against this atmospheric solution. 

To investigate the plausibility of the retrieved \ce{SO}, we performed additional photochemical simulations using VULCAN \citep{tsai2021}. We adopted the isothermal temperature profile from Model I and assumed a uniform eddy diffusion coefficient of $K_{zz} = 10^8 cm^2 s^{-1}$ for vertical mixing. As a proxy for NGTS-2, we used a semi-empirical stellar spectrum with $T_{\mathrm{eff}}$ = 6500 K from \cite{rugheimer2013}. As shown in Figure \ref{fig:spaghetti_vulcan}, we find that as the temperature decreases below approximately 700 K to the retrieved temperature of Model I, \ce{SO2} starts to decline due to the lack of OH radicals \citep{tsai2023, crossfield2025}. Instead, SO becomes more abundant than \ce{SO2}. However, the abundance of SO remains below roughly $10^{-6}$, lower than the detection limit and our retrieved SO abundance. Therefore, the retrieved temperature and SO abundances in Model I are likely not physically consistent. A full exploration of sulphur photochemistry expectations of NGTS-2b, along with the full BOWIE-ALIGN sample, will be undertaken in a follow-up paper. 

We additionally assess why our observations support this atmospheric solution, we perform a leave-one-out cross validation analysis on \poseidon{} Model I (with \ce{SO}) and Model II (without \ce{SO}). The analysis is described in Appendix \ref{app:loocv}, following the methods outlined in \citet{welbanks2023loocv}. Figure \ref{fig:loocv} shows our cross validation results on the \poseidon{} $R$\,=\,100 spectrum, where the two data points at 4.29 and 4.33\,$\upmu$m are responsible for the high \ce{SO} abundance due to their low transit depths on the blue edge of the \ce{CO2} feature. These points are consistent between both reductions, therefore the detection is unlikely to be due to reduction level choices.
Decomposing the retrieved spectrum of Model I (Figure \ref{fig:loocv}) shows how the 4--5\um spectral feature is fit by the combined opacity of \ce{SO} at the  blue edge, and \ce{CO2} at the red edge. Fitting the spectral feature in this way can only be accomplished with a relatively low abundance of \ce{CO2} to a high abundance of \ce{SO}, but most critically, with low atmospheric temperatures. Increasing the temperature from 630\,K to 1500\,K quickly leads to the \ce{CO2} opacity dominating the \ce{SO} opacity, with both spectral features becoming broader, such that their combined opacity does not fit the data. Given that such low limb temperatures are not expected for hot Jupiters with equilibrium temperatures around 1500 K \citep[e.g.][]{kataria2016}, even if significant quantities of \ce{SO} are present in the atmosphere of NGTS-2b, it is unlikely that \ce{SO} opacity would be observable in the NIRSpec/G395H bandpass. 

\subsubsection{Excess absorption at 4.8 \um}
We note that a rise in the transmission spectrum occurs between 4.8 -- 4.9\um, which may be indicative of an absorption signature. Compared to the models we have tested, this rise in transit depths remains as excess absorption, which is not fit by the gaseous species within the models. WASP-15b showed a similar feature at 4.9 \um \citep{kirk2025_w15}, for which \ce{OCS} was considered to be a leading candidate, with a number of caveats: the width of the \ce{OCS} feature was broader than the observed transmission signature, chemical networks do not expect significant abundances of \ce{OCS} at the pressures probed in transmission, and \ce{OCS} is photochemically depleted \citep{tsai2021, tsai2023}, such that its presence is unlikely to be observed in the atmospheres of hot Jupiters around F type stars. 
The spectra from both NGTS-2b and WASP-15b are binned to the same wavelength grid, showing that, from the $R=100$ spectrum, the excess absorption is centred on adjacent wavelength points (4.83 \um for NGTS-2b and 4.88 \um for WASP-15b). 
We run an additional retrieval on the \poseidon{}  $R=400$ spectrum, where the model parameters are identical to Model II with the inclusion of \ce{OCS}. From this retrieval, we do not find that absorption of \ce{OCS} is statistically favoured ($\ln{Z} = 1798.73$). The abundance of \ce{OCS} is retrieved unconstrained, with a $3\sigma$ VMR upper limit of -2.99 dex.

\subsection{High mean molecular weight modes}\label{sec:mmw}

\begin{figure}
    \centering
    \includegraphics[width=\linewidth]{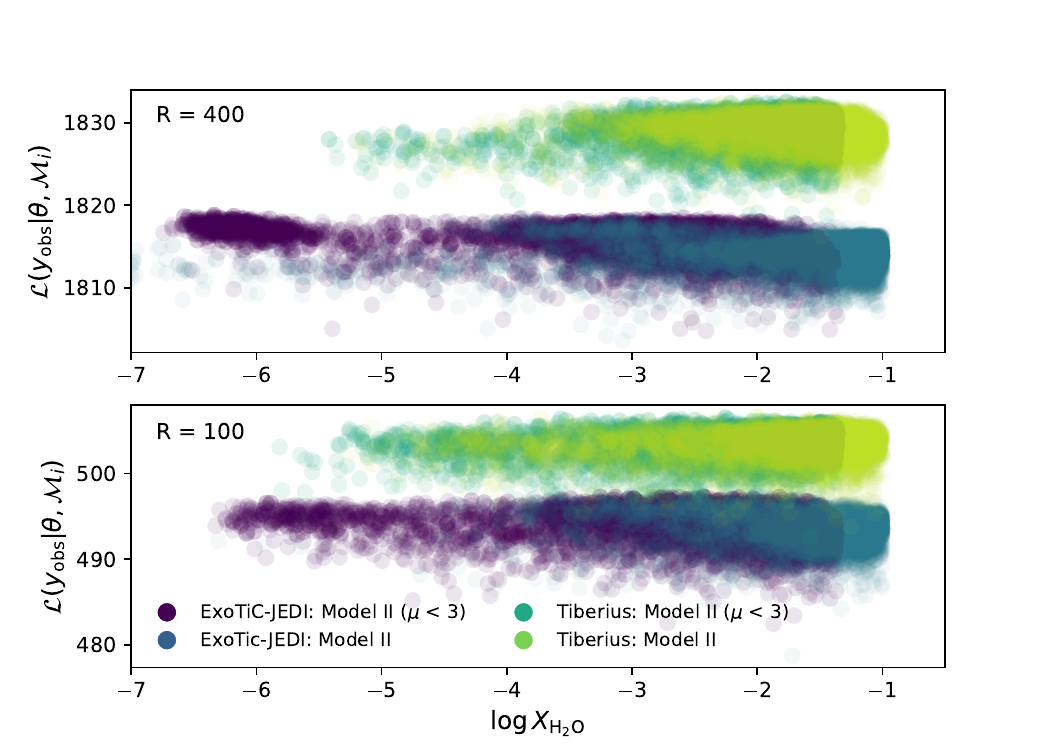}
    \caption{Likelihood samples for \poseidon{} Model II and II ($\mu < 3$) demonstrating the effect of the prior volume on the resultant posterior distribution. When we restrict the upper mean molecular weight limit to 3, the distribution is equally flat but truncated by the prior volume. This indicates that in both model set-ups the results are uninformed by the data. }
    \label{fig:loglike}
\end{figure}

The high mean molecular weights due to high \ce{H2O} VMRs found in \poseidon{} Model II across the reductions are not uncommon in retrieval analyses of hot Jupiter atmospheres \citep{wakeford2018, lueber2024informationcontent,  kirk2025_w15}, and the impact of mean molecular weight (or by proxy, metallicity) on scale height is well established \citep[e.g.,][]{goyal2019}. 
When performing atmospheric inference with retrievals, the set of parameters that make up our model atmosphere ultimately influence the transit depth of the model spectrum. For the case of species abundances (in particular, \ce{H2O}, due to its high abundance and dominant spectral features in hot Jupiter atmospheres), the effect on transit depth is twofold. At low VMRs, the dominant effect is the strength of the absorption feature whereas at high VMRs, muting across all wavelengths occurs due to the mean molecular weight.
We see that high mean molecular weights dominate our posterior distributions in Model II (Figure \ref{fig:mmw_hist}). However, by definition from Bayes theorem, where the posterior is the product of the likelihood and prior normalized to the evidence, we must consider the contribution of both the model likelihood and prior. When discussing the prior, we refer to the full phase space being explored by all variable parameters, which ultimately control the transit depth, as the total prior volume. In plotting the likelihood of samples against \ce{H2O} abundance (Figure \ref{fig:loglike}), we see a flat distribution, indicating a more degenerate parameter space than is implied by the posteriors. Therefore, it must be the prior volume that is biasing the retrieved \ce{H2O} and \ce{CO2} abundance, given the data are uninformative. At high VMRs, changing the \ce{H2O} abundance has a smaller impact on the transit depth than at lower VMRs. As such, the total prior volume within this region is larger, leading to the oversampling of this region of parameter space, despite the use of the same uniform log-normal abundance priors. 
This leads to two main conclusions from our free chemistry retrievals. First, the NIRSpec/G395H observations, through a combination of precision, wavelength coverage and random scatter, are not sufficiently informative to confidently constrain the atmosphere. Second, we cannot rule out low probability solutions from our Model II retrieval results.

\subsubsection{Cloud-Metallicity degeneracy}
Our equilibrium chemistry retrievals suggest that a lower metallicity than that inferred from free chemistry can fit the observations when a grey cloud opacity is present. 
The cloud-mean molecular weight degeneracy is well documented in the literature \citep[e.g.,][]{bennekeseager2013, knutson2014}, and \citet{lineparmentier2016} showed analytically, how cloud-free, high mean molecular weight atmospheres can mimic solar composition atmospheres with inhomogeneous cloud cover.  
There have been many studies investigating biases in one-dimensional retrievals of synthetic multidimensional atmospheres, showing that temperature contrasts \citep[e.g.,][]{caldas2019, pluriel2020} and aerosol coverage \citep[e.g.,][]{lacyburrows2020, macdonald2020} can bias the retrieved solution. 
Recent observational evidence of asymmetric terminators with cooler, cloudy morning limbs and hotter, cloud free evening limbs has emerged \citep[e.g.,][]{espinoza2024, murphy2024, murphy2025, muckherjee2025}, whereby biased metallicity from the terminator averaged spectrum has been inferred by NIRISS/SOSS observations of WASP-94Ab \citep{muckherjee2025}.
The exploration of potential asymmetries biasing our retrieval results through the extraction of morning and evening spectra will be covered in a follow up study on the full BOWIE-ALIGN sample. However, we do explore the effect of inhomogeneous cloud opacity by performing an additional test on Model II, adding a patchy cloud parameter, $\phi_\mathrm{cloud}$, that encodes the fraction of the terminator where cloud opacity is present \citep{macdonaldmadhusudhan2017poseidon, macdonald2020}. 

\begin{figure}
    \centering
    \includegraphics[width=\linewidth]{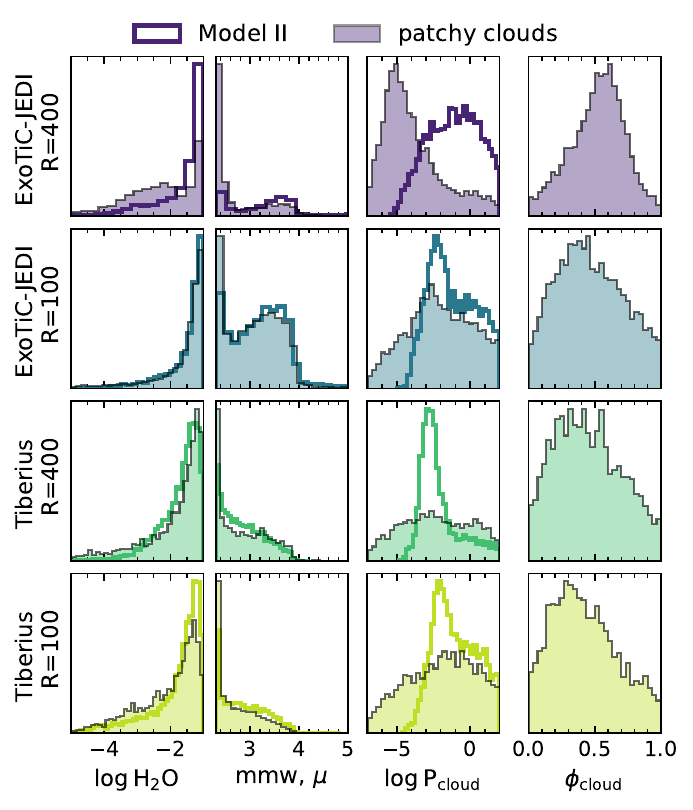}
    \caption{Probability density histograms comparing the inclusion of patchy clouds on the Model II \poseidon{} retrievals across our two resolutions and two reductions. The inclusion of patchy clouds only affects the retrieved \ce{H2O} volume mixing ratio for the \jedi{} $R$=400 reduction and is not statistically favoured over homogeneous cloud coverage.}
    \label{fig:patchy}
\end{figure}

We run an inhomogeneous cloud test with \poseidon{} on all four reductions. For the \tiberius{} reductions and \jedi{} $R$=400 reduction, adding patchy clouds leads to \ce{H2O} and mean molecular weights consistent with Model II, although constraints on the cloud pressure level worsen (Figure \ref{fig:patchy}). In contrast, the \jedi{} $R$=400 reduction retrieves a high altitude cloud deck ($\log{P_\mathrm{cloud}} = -4.42^{+2.56}_{-1.28}$) with a coverage fraction of $0.54^{+0.35}_{-0.45}$. With this cloud state, a lower \ce{H2O} abundance of $\log{\mathrm{H_2O}} = -2.24^{+1.07}_{-1.22}$, although a high mean molecular weight mode is still retrieved. These solutions have a marginally higher evidence than Model II for the \jedi{} reductions and lower evidence for \tiberius{}. From this, there is insufficient evidence to support the presence of patchy clouds, and even in the \jedi{} $R$=400 retrieval, the coverage fraction is poorly constrained. 
\\

\noindent From our presented measurements 
it is clear that the \ce{H2O} abundance and cloud opacity cannot be well constrained together from NIRSpec/G395H alone. Whilst NIRSpec/G395H is crucial in understanding the carbon content of atmospheres \citep[e.g,][]{alderson2023g395h_w39ers}, the 3--5\,$\upmu$m range is least impacted by cloud opacity and provides coverage of only half an \ce{H2O} absorption feature. As such, in the absence of strong absorption signatures, additional wavelength coverage is often necessary to confidently constrain the atmospheric composition \citep[e.g.,][]{verma2025} and break such cloud-abundance degeneracies. 

NGTS-2b has since been observed with NIRISS/SOSS in Cycle 3 (GO 5924, PI: Sing), and from 0.38--0.7\,\um with the ground-based NTT/EFOSC2 instrument (PI Ahrer). These observations can provide additional constraints on the \ce{H2O} abundance, with NIRISS/SOSS covering multiple absorption bands \citep{fisher2024}, and NTT data \citep[e.g.,][]{ahrer2022ntt, ahrer2023ntt} proving additional short wavelength information crucial in constraining cloud opacity \citep{fairman2024}.

\subsection{Implications for planet formation}
In our analysis, we find evidence for a stellar to super-stellar atmospheric metallicity and a weak constraint of the C/O ratio of the planet's atmosphere between 0.2 and 0.8. The planet's composition is therefore naturally reproduced by a wide range of formation scenarios (and locations) in which the envelope is enriched by solids. These solids could have been, e.g., delivered during formation by planetesimals or pebbles \citep{Mordasini2016,Booth2017,Madhusudhan2017,Penzlin2024_BOWIE}, mixed out from the core \citep{Knierim2024}, or delivered late, e.g., by comets \citep{Sainsbury-Martinez2024}. Since sub-stellar metalliticies are disfavoured, we can rule out formation scenarios in which NGTS-2~b's envelope composition is dominated by gas accreted in a disc with sub-solar metallicity gas \citep[e.g., some of the scenarios discussed in][]{Penzlin2024_BOWIE}. Such sub-stellar gas metallicities are frequently inferred from ALMA observations of the outer regions of protoplanetary discs (specifically, C/H is often depleted by a factor 10 together with C/O $\gtrsim 1$, as discussed in \citealt{Bergin2024}) and slightly sub-solar gas metallicities are common in planet formation models that are based on \citet{Oberg2011}. However, models with efficient pebble drift could explain the high metallicity of NGTS-2b via the accretion of gas enriched by volatiles sublimating off of drifting pebbles interior to the regions constrained by ALMA \citep[e.g.,][]{Booth2017,Schneider_Bitsch2021,Danti2023}, thus enrichment by solids is possible but not essential. Even with these relatively weak constraints, the results can help strengthen statistical arguments together with larger samples of planetary compositions, e.g., as in the BOWIE-ALIGN programme \citep{Kirk2024}.

\section{Conclusions}\label{sec:conclusions}

This study presents the first atmospheric observations and characterisation of NGTS-2b. Using JWST NIRSpec/G395H, we measure the transmission from 2.84--5.18\,$\upmu$m using two different reduction pipelines, \jedi{} and \tiberius{}. These pipelines produce consistent results well within one-sigma for both R\,=\,100 and R\,=\,400 reductions. To interpret the NGTS-2b transmission spectrum, we use the \poseidon{} and \prt{} retrieval suites, finding generally consistent results across model frameworks, given the uncertainty in our retrieved parameters. 

We find that the atmosphere shows absorption features that predominantly correspond to \ce{H2O} and \ce{CO2}, however, we are unable to place meaningful constraints on their abundances, likely due to known cloud-metallicity degeneracies that cannot be resolved in this wavelength range given the SNR of our data. 
Our results converge on high mean molecular weight solutions, where absorption features are fit with high \ce{H2O} abundances influencing the scale height of the atmosphere. We show that such solutions are not predicated on higher likelihood regions of the parameter space and are therefore mainly influenced by the prior. In cases such as NGTS-2b, where the observations are generally uninformative, it is therefore important to exercise caution when interpreting retrieved posterior distributions. 
Whilst present across all reductions and retrievals, this high mean molecular weight result is amplified in the \poseidon{} retrieval on the \jedi{} $R$\,=\,400 reduction, likely a combination of both reduction choices and retrieval parameterisation.

Using interior structure models, we find that atmospheres with metallicities greater than 43.5$\times$ solar are not expected. Placing a restriction on the mean molecular weight of our retrieval, we find more astrophysically plausible solutions. Nevertheless, we are still limited in our interpretation of the planet beyond \ce{H2O} and \ce{CO2} identification due to our priors restricting the posterior distributions, which maintain the same underlying distribution as when no restriction is placed on the mean molecular weight. 

The data shows hints of favouring solutions with the presence of SO at a significance of 2$\sigma$. However, we show that the abundances retrieved are likely not physically feasible, given the recovered metallicity and temperature of the atmosphere. We demonstrate that the presence of \ce{SO} is driven by two datapoints in the blue end of the \ce{CO2} feature through leave-one-our cross validation, and that this fit can only be recovered at implausibly low atmospheric temperatures. 

We are unable to place strong constraints on the atmospheric metallicity and C/O ratio through our equilibrium chemistry retrievals. Although all retrievals support a supersolar atmosphere, solar metallicity cannot be ruled out by the \prt{} retrievals. With the exception of the \poseidon{} retrieval on the \jedi{} $R$\,=\,400 spectrum, our results show a weak preference for lower C/O ratios. Furthermore, our \poseidon{} retrievals support the presence of a high altitude cloud deck despite \prt{} producing unconstrained posteriors. Therefore, our equilibrium retrievals do not allow us to further interpret the role of cloud opacity in the NIRSpec/G395H spectrum of NGTS-2b. With the majority of retrievals ruling out sub-solar metallicity, our results disfavour gas accretion formation scenarios from a metal depleted disc. 

Future studies as part of BOWIE-ALIGN looking at the role of photochemistry and limb-asymmetries may further elucidate our findings for the atmosphere of NGTS-2b. In addition, observations expanding the wavelength coverage explored in the atmosphere may also aid in future interpretation.  These observations represent one of four planets in the BOWIE-ALIGN sample for aligned planetary systems. While the large uncertainties in NGTS-2~b's composition mean that the constraints on its formation are weak, the results can help strengthen statistical arguments about the origins of the hot Jupiter population more generally, as will be explored in the upcoming BOWIE-ALIGN population papers.

\section*{Acknowledgements}
We thank Ryan MacDonald for helpful discussions on the \poseidon{} retrieval framework.
This work is based on observations made with the NASA/ESA/CSA JWST. The data were obtained from the Mikulski Archive for Space Telescopes at the Space Telescope Science Institute, which is operated by the Association of Universities for Research in Astronomy, Inc., under NASA contract NAS 5-03127 for JWST. These observations are associated with program \#3838. This work was inspired by collaboration through the UK-led BOWIE+ collaboration. Support for program JWST-GO-3838 was provided by NASA
through a grant from the Space Telescope Science Institute, which is operated by the
Association of Universities for Research in Astronomy, Inc., under NASA contract NAS
5-03127. 

C.F is funded by the University of Bristol PhD scholarship fund. 
H.R.W was funded by UK Research and Innovation (UKRI) under the UK government’s Horizon Europe funding guarantee as part of an ERC Starter Grant [grant number EP/Y006313/1].
J.K acknowledges financial support from Imperial College London through an Imperial College Research Fellowship grant. 
R.A.B thanks the Royal Society for their support through a University Research Fellowship and the SFTC via award UKRI1191.
N.J.M. and M.Z. acknowledge support from a UKRI Future Leaders Fellowship [Grant MR/T040866/1], a Science and Technology Facilities Funding Council Nucleus Award [Grant ST/T000082/1], and the Leverhulme Trust through a research project grant [RPG-2020-82].
P.J.W.\ acknowledges support from the UK Science and Technology Facilities Council (STFC) through consolidated grant ST/X001121/1.

\section*{Data Availability}
The raw data are available on the Mikulski Archive for Space Telescopes at the Space Telescope Science Institute, under program number \#3838. The data products associated with this manuscript, including spectra and retrieval outputs are available on Zenodo, via https://doi.org/10.5281/zenodo.18864528.
 



\bibliographystyle{mnras}
\bibliography{bibliography} 




\appendix

\section{Leave-one-out cross validation}\label{app:loocv}
\begin{figure}
    \centering
    \includegraphics[width=\linewidth]{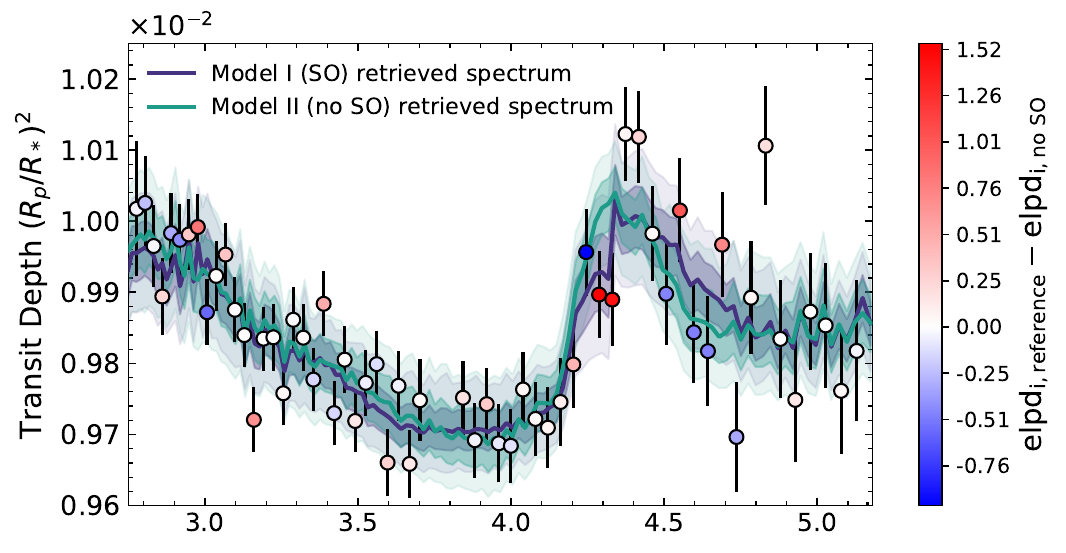}
    \includegraphics[width=\linewidth]{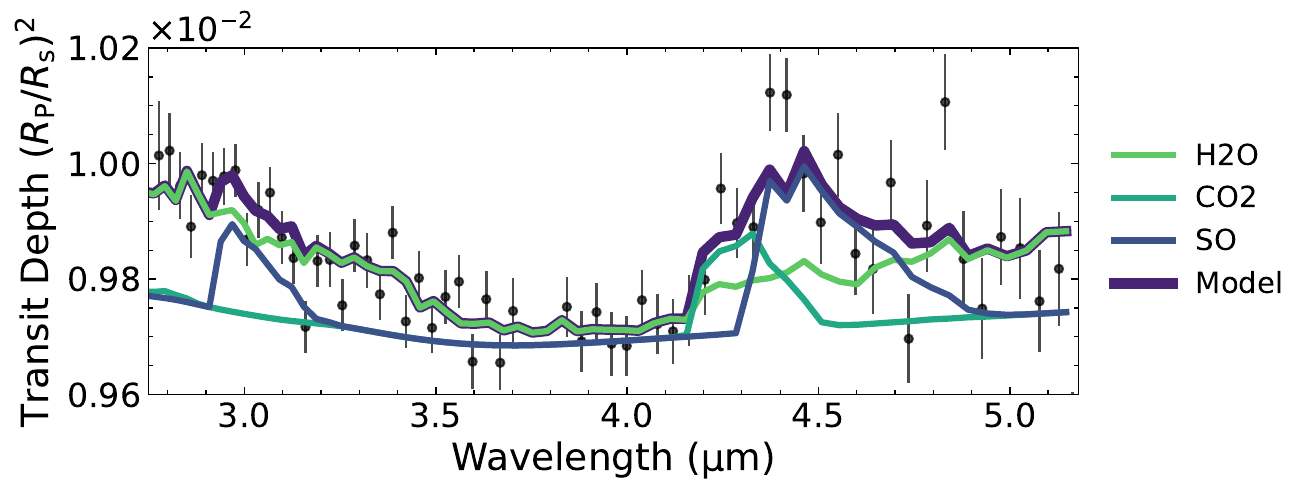} 
    \includegraphics[width=\linewidth]{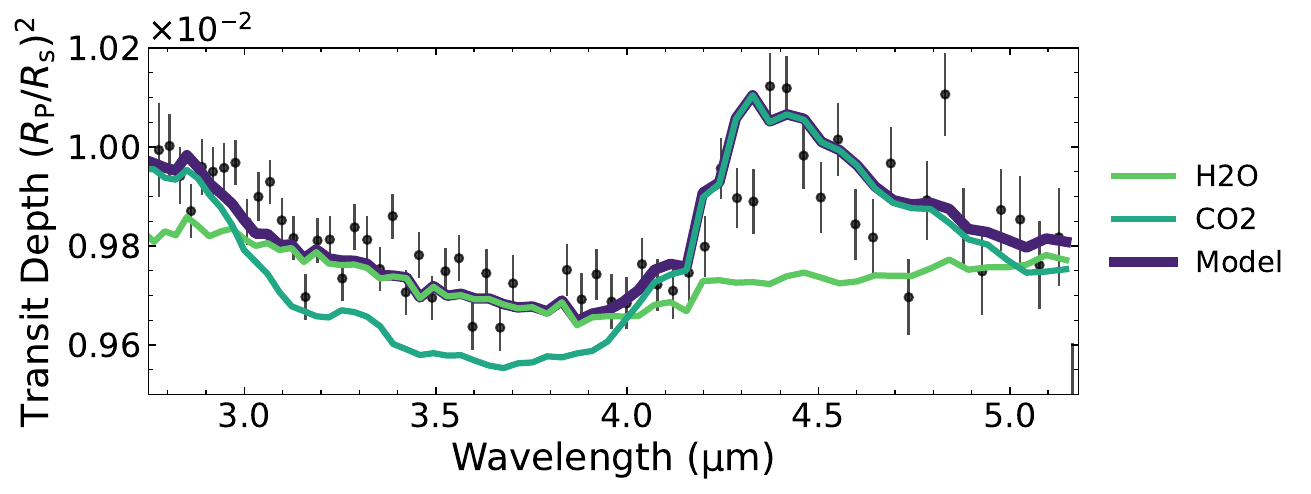}
    \caption{Leave-one-out cross-validation transmission spectral analysis on the \jedi{} $R$\,=\,400 spectrum of NGTS-2b. Bolder colours represent larger contributions to the comparative spectral models (with and without SO). This shows that the inclusion of SO in the model is driven predominantly by two/three datapoints on the left side of the \ce{CO2} feature at around 4.25\,$\upmu$m. The middle figure shows the spectral contributions for each of the main species considered in Model I (including SO) where the position of the 'shoulder' between the \ce{CO2} and SO absorption features meet corresponds to the datapoints driving the fit statistics. The bottom figure shows the spectral contributions for \ce{H2O} and \ce{CO2} for Model II, no other species or cloud opacity show significant contributions to the spectrum.}
    \label{fig:loocv}
\end{figure}

To assess the influence of individual data point contribution on the retrieved \ce{SO} abundance, we perform a leave-one-out cross validation analysis as described in \citep{welbanks2023loocv}. This method calculates the expected log posterior predictive density for the $i^{th}$ datapoint in a transmission spectrum,

\begin{equation}
    elpd_{loo, i, \mathcal{M}} = \log{p(D_i|D_{-i}, \mathcal{M})}
\end{equation}

quantifying how well the the dataset minus the $i^{th}$ datapoint, $D_{-i}$, can predict the missing data. To reduce the computation to a single retrieval for each model, we implement the Pareto Smoothed Importance Sampling (PSIS) approximation \citep{Vehtari2016} on Model I (with \ce{SO}) and Model II (without \ce{SO}), for the \jedi{} $R$\,=\,100 reduction, with Model I acting as the fiducial model. Our Pareto-$k$ values are all $<0.7$, the empirical limit at which the PSIS approximation is likely to be reliable \citep{Vehtari2016, welbanks2023loocv}. Figure \ref{fig:loocv} shows the pointwise difference in $elpd_{loo, i, \mathcal{M}}$ between Model I and II, where higher values indicate datapoints that are better predicted by atmospheres including \ce{SO} opacity than atmospheres excluding \ce{SO} opacity.

\section{Retrieval results}\label{app:retrieval_results}


\renewcommand{\arraystretch}{1.3}
\setlength{\tabcolsep}{3pt}
\begin{table*}
    \caption{Atmospheric retrieval results: The retrieved median species abundances and 1$\sigma$ errors for our \poseidon{} and \prt{} free chemistry retrievals. All values are displayed are volume mixing ratios (VMRs), where the \prt{} abundances have been converted from mass fractions using the median mean molecular weight for each model. For the models with SO (\textbf{Model A}), these values are 3.33 (\jedi{} $R$=400), 3.15 (\jedi{} $R$=100), 3.37 (\tiberius{} $R$=400), and 3.62 (\tiberius{} $R$=100). For the models without SO (\textbf{Model B}), these values are 2.81 (\jedi{} $R$=400) and 2.95 (\tiberius{} $R$=400).}
    \label{tab:retrieval_results_free_abundances}
    \centering
    \begin{tabular}{l c c c c c c c c c c c } 
    \hline
     Observation + Retrieval  & $\log{X_\mathrm{H_2O}}$ & $\log{X_\mathrm{CO}}$   & $\log{X_\mathrm{CO_2}}$  & $\log{X_\mathrm{CH_4}}$  & $\log{X_\mathrm{NH_3}}$  & $\log{X_\mathrm{HCN}}$  & $\log{X_\mathrm{H_2S}}$  & $\log{X_\mathrm{SO_2}}$  & $\log{X_\mathrm{SO}}$  \\ 
     \hline
     \textbf{\jedi{} $R$=400 } & & & & & & & \\
     \poseidon{}: \textit{free - Model I (SO)} & $-3.51^{+0.73}_{-0.67}$ &  $-8.06^{+2.51}_{-2.66}$ &  $-7.38^{+0.94}_{-0.81}$ &  $-9.83^{+1.58}_{-1.50}$ &  $-9.23^{+1.92}_{-1.86}$ &  $-9.30^{+1.77}_{-1.82}$ &  $-8.70^{+2.40}_{-2.19}$ &  $-9.88^{+1.52}_{-1.41}$ &  $-2.73^{+0.57}_{-0.77}$ \\
     \poseidon{}: \textit{ free - Model II } & $-1.17^{+0.11}_{-0.23}$ &  $-6.23^{+2.63}_{-3.43}$ &  $-4.32^{+0.33}_{-0.46}$ &  $-7.32^{+1.83}_{-3.07}$ &  $-7.71^{+2.45}_{-2.66}$ &  $-8.27^{+2.06}_{-2.21}$ &  $-7.32^{+2.73}_{-2.82}$ &  $-8.97^{+1.88}_{-1.86}$ & -- \\
     \poseidon{}: \textit{free - Model II ($\mu < 3$) } & $-1.86^{+0.36}_{-0.73}$ &  $-6.15^{+2.37}_{-3.47}$ &  $-4.78^{+0.42}_{-0.66}$ &  $-7.04^{+1.80}_{-2.88}$ &  $-8.57^{+2.50}_{-2.17}$ &  $-8.54^{+2.12}_{-2.12}$ &  $-7.02^{+2.84}_{-3.10}$ &  $-8.81^{+2.01}_{-1.93}$ & -- \\
     \\
    \texttt{pRT}: \textit{free - Model A (SO)} &
     $-3.30^{+0.91}_{-0.76}$ &  $-7.81^{+3.03}_{-3.51}$ &  $-7.04^{+1.29}_{-0.91}$ &  $-9.89^{+1.83}_{-1.94}$ &  -- &  -- &  $-8.80^{+2.80}_{-2.84}$ &  $-10.34^{+1.97}_{-1.97}$ & $-2.90^{+0.72}_{-0.98}$ \\
     \texttt{pRT}: \textit{ free - Model B } & 
     $-2.23^{+0.61}_{-0.93}$ &  $-3.50^{+1.45}_{-5.08}$ &  $-4.90^{+0.72}_{-0.89}$ &  $-9.34^{+2.23}_{-2.22}$ &  -- &  -- &  $-8.46^{+3.04}_{-2.96}$ &  $-9.62^{+2.60}_{-2.36}$ & -- \\
     \texttt{pRT}: \textit{ hybrid} & 
     -- &  -- &  -- &  -- &  -- &  -- &  $-8.11^{+3.15}_{-3.20}$ &  $-9.60^{+2.50}_{-2.40}$ & $-6.51^{+3.23}_{-4.30}$ \\
     \hline 
     \textbf{\jedi{} $R$=100 } & & & & & & & \\
     \poseidon{}: \textit{free - Model I (SO)} & $-3.71^{+0.84}_{-0.80}$ &  $-8.34^{+2.42}_{-2.50}$ &  $-7.18^{+1.06}_{-0.75}$ &  $-9.90^{+1.50}_{-1.42}$ &  $-9.36^{+1.92}_{-1.78}$ &  $-9.24^{+1.92}_{-1.86}$ &  $-8.65^{+2.38}_{-2.25}$ &  $-9.74^{+1.59}_{-1.51}$ &  $-2.91^{+0.67}_{-0.91}$ \\
     \poseidon{}: \textit{ free - Model II } & $-1.28^{+0.21}_{-0.78}$ &  $-4.93^{+2.46}_{-4.38}$ &  $-3.85^{+0.75}_{-0.91}$ &  $-9.01^{+1.95}_{-1.88}$ &  $-8.22^{+2.54}_{-2.46}$ &  $-8.08^{+2.38}_{-2.42}$ &  $-7.55^{+2.87}_{-2.82}$ &  $-8.76^{+1.94}_{-2.00}$ & -- \\
     \poseidon{}: \textit{free - Model II ($\mu < 3$) } & $-2.30^{+0.65}_{-1.16}$ &  $-3.87^{+1.36}_{-3.76}$ &  $-4.92^{+0.69}_{-1.11}$ &  $-9.09^{+1.90}_{-1.83}$ &  $-8.73^{+2.20}_{-2.06}$ &  $-8.81^{+2.20}_{-2.06}$ &  $-7.87^{+2.83}_{-2.62}$ &  $-8.91^{+1.99}_{-1.95}$ & -- \\
     \\
    \texttt{pRT}: \textit{free - Model A (SO)} &
      $-3.23^{+0.82}_{-0.94}$ &  $-8.29^{+2.87}_{-3.00}$ &  $-6.67^{+1.06}_{-0.89}$ &  $-9.44^{+1.88}_{-2.26}$ &  -- &  -- &  $-8.70^{+2.86}_{-2.81}$ &  $-10.21^{+2.06}_{-2.04}$ & $-2.90^{+0.75}_{-1.04}$ \\
     \texttt{pRT}: \textit{ hybrid} &
      -- &  -- &  -- &  -- &  -- &  -- &  $-8.42^{+3.44}_{-3.08}$ &  $-9.57^{+2.45}_{-2.49}$ & $-5.46^{+2.56}_{-4.99}$ \\

     \hline 
     \textbf{\tiberius{} $R$=400 } & & & & & & & \\
     \poseidon{}: \textit{free - Model I (SO)} & $-2.54^{+1.25}_{-1.41}$ &  $-8.22^{+2.80}_{-2.49}$ &  $-6.63^{+2.06}_{-1.31}$ &  $-8.91^{+1.75}_{-2.02}$ &  $-8.88^{+2.14}_{-2.05}$ &  $-8.86^{+1.95}_{-2.01}$ &  $-7.84^{+2.79}_{-2.74}$ &  $-9.47^{+1.76}_{-1.68}$ &  $-3.48^{+1.02}_{-2.49}$  \\
     \poseidon{}: \textit{ free - Model II } & $-1.55^{+0.36}_{-0.76}$ &  $-6.81^{+3.45}_{-3.32}$ &  $-4.67^{+0.54}_{-0.73}$ &  $-8.33^{+2.07}_{-2.36}$ &  $-8.63^{+2.43}_{-2.18}$ &  $-8.73^{+2.07}_{-2.12}$ &  $-7.75^{+2.80}_{-2.70}$ &  $-8.91^{+2.06}_{-1.88}$ & -- \\
     \poseidon{}: \textit{free - Model II ($\mu < 3$) } & $-1.93^{+0.37}_{-0.66}$ &  $-6.16^{+2.89}_{-3.70}$ &  $-5.01^{+0.51}_{-0.67}$ &  $-7.97^{+1.77}_{-2.55}$ &  $-8.66^{+2.25}_{-2.10}$ &  $-8.96^{+2.09}_{-1.97}$ &  $-7.92^{+2.86}_{-2.59}$ &  $-8.89^{+2.04}_{-1.96}$ & -- \\
     \\
    \texttt{pRT}: \textit{free - Model A (SO)} & 
     $-3.02^{+0.78}_{-0.79}$ &  $-8.66^{+2.75}_{-2.96}$ &  $-7.35^{+1.14}_{-1.33}$ &  $-9.44^{+1.88}_{-2.26}$ &  -- &  -- &  $-8.97^{+2.71}_{-2.75}$ &  $-10.36^{+1.94}_{-1.98}$ & $-2.85^{+0.70}_{-0.85}$ \\
     \texttt{pRT}: \textit{ free - Model II } & 
     $-2.12^{+0.56}_{-0.87}$ &  $-4.48^{+2.09}_{-5.19}$ &  $-5.07^{+0.71}_{-0.87}$ &  $-8.49^{+2.09}_{-2.71}$ &  -- &  -- &  $-8.76^{+3.19}_{-2.86}$ &  $-9.76^{+2.46}_{-2.32}$ & -- \\
     \texttt{pRT}: \textit{ hybrid} &
      -- &  -- &  -- &  -- &  -- &  -- &  $-8.56^{+3.02}_{-3.01}$ &  $-9.55^{+2.47}_{-2.53}$ & $-6.61^{+2.81}_{-4.41}$ \\

     \hline 
     \textbf{\tiberius{}{} $R$=100 } & & & & & & & \\
     \poseidon{}: \textit{free - Model I (SO)} & $-3.28^{+0.86}_{-0.83}$ &  $-8.59^{+2.47}_{-2.27}$ &  $-7.18^{+1.06}_{-1.00}$ &  $-9.66^{+1.62}_{-1.59}$ &  $-9.06^{+1.97}_{-1.97}$ &  $-8.96^{+1.93}_{-1.99}$ &  $-8.46^{+2.51}_{-2.42}$ &  $-9.62^{+1.61}_{-1.56}$ &  $-2.89^{+0.65}_{-0.92}$  \\
     \poseidon{}: \textit{ free - Model II } & $-1.53^{+0.35}_{-1.09}$ &  $-6.63^{+2.88}_{-3.42}$ &  $-4.73^{+0.66}_{-1.12}$ &  $-9.10^{+1.92}_{-1.85}$ &  $-8.40^{+2.27}_{-2.24}$ &  $-8.50^{+2.32}_{-2.16}$ &  $-7.71^{+2.75}_{-2.69}$ &  $-9.03^{+1.91}_{-1.91}$ & -- \\
     \poseidon{}: \textit{free - Model II ($\mu < 3$) } & $-2.05^{+0.48}_{-1.00}$ &  $-6.26^{+2.67}_{-3.57}$ &  $-5.21^{+0.63}_{-0.99}$ &  $-9.05^{+1.90}_{-1.85}$ &  $-8.63^{+2.22}_{-2.15}$ &  $-8.72^{+2.12}_{-2.03}$ &  $-7.78^{+2.68}_{-2.63}$ &  $-9.05^{+1.82}_{-1.86}$ & -- \\
     \\
    \texttt{pRT}: \textit{free - Model A (SO)} & 
     $-2.91^{+0.77}_{-0.75}$ &  $-8.81^{+2.80}_{-2.62}$ &  $-6.91^{+0.96}_{-1.09}$ &  $-9.74^{+1.82}_{-1.87}$ &  -- &  -- &  $-8.87^{+2.82}_{-2.67}$ &  $-10.11^{+1.94}_{-2.05}$ & $-2.67^{+0.66}_{-0.86}$ \\
     \texttt{pRT}: \textit{ hybrid} &
      -- &  -- &  -- &  -- &  -- &  -- &  $-8.64^{+3.07}_{-2.93}$ &  $-9.67^{+2.52}_{-2.44}$ & $-5.20^{+2.21}_{-5.13}$ \\

    \end{tabular}

\end{table*}
\setlength{\tabcolsep}{1pt}

\renewcommand{\arraystretch}{1.1}
\setlength{\tabcolsep}{3pt}
\begin{table*}
    \caption{Atmospheric retrieval results continued: Retrieved parameters and statistics for our \poseidon{} and \prt{} free and equilibrium chemistry retrievals. For model comparison, we include the model evidence  $\ln{\mathcal{Z}}$, $\chi^2_{\nu}$ and degrees of freedom (dof).}
    \label{tab:retrieval_results_main}
    \centering
    \begin{tabular}{l c c c c c c c c c c c} 
    \hline
     Observation + Retrieval & $M/H$ ($\times$ solar) & C/O & $R_{\mathrm{P_{ref}}}$ & $\log{g}$ & $T$ (K) & $\log{\mathrm{P_{cloud}}}$ (bar) & $\delta_\mathrm{rel}$ (ppm) & $\ln{\mathcal{Z}}$ & $\chi^2_{\nu}$ & dof \\
     \hline
     \textbf{\jedi{} $R$=400 } & & & & & & & \\
     \poseidon{}: \textit{free - Model I (SO)} & - & - & $1.615^{+0.004}_{-0.005}$  &  $2.85^{+0.02}_{-0.02}$ &  $632^{+93}_{-65}$ &  - & $60^{+20}_{-21}$ & 1802.12 & 1.18 & 226 \\
     \poseidon{}: \textit{ free - Model II } & - & - & $1.600^{+0.010}_{-0.030}$ &  $2.87^{+0.02}_{-0.02}$ &  $916^{+236}_{-131}$ &  $-1.11^{+1.84}_{-1.97}$ &  $84^{+42}_{-41}$ & 1799.04 & 1.20 & 226\\
     \poseidon{}: \textit{free - Model II ($\mu < 3$) } & - & - & $1.530^{+0.040}_{-0.050}$ &  $2.86^{+0.02}_{-0.02}$ &  $1507^{+379}_{-350}$ &  $-3.63^{+2.72}_{-0.92}$ &  $72^{+24}_{-23}$ & 1799.20 & 1.20 & 226  \\
     \poseidon{}: \textit{equilibrium} & $12.88^{+26.30}_{-5.25}$ & $0.61^{+0.18}_{-0.23}$ & $1.55^{+0.02}_{-0.02}$ & $2.86^{+0.02}_{-0.02}$ & $1340^{+191}_{-219}$ & $-3.53^{+0.83}_{-0.62}$ & $73^{+18}_{-19}$ & 1802.97 & 1.17 & 232\\
     \\
    \texttt{pRT}: \textit{free - Model I (SO)} & - & - & 
    $1.697\pm0.008$ &  $2.81\pm0.07$ &  $637^{+189}_{-89}$ & $-0.09^{+1.40}_{-1.45}$ &  $58\pm20$  & 1801.1 & 1.15 & 227 \\
    \texttt{pRT}: \textit{ free - Model II } & - & - &
    $1.677\pm0.012$ &  $2.89^{+0.05}_{-0.06}$ &  $1021^{+193}_{-155}$ & $-2.61^{+0.96}_{-0.78}$ &  $56\pm22$  & 1800.5 & 1.39 & 228 \\
    \texttt{pRT}: \textit{ hybrid} & $5.13^{+22.41}_{-3.62}$ & $0.35^{+0.18}_{-0.15}$ &
    $1.682\pm0.006$ &  $2.87^{+0.06}_{-0.07}$ &  $987^{+113}_{-71}$ & $-2.26^{+2.12}_{-0.63}$ &  $59\pm22$  & 1809.7 & 1.27 & 229 \\
    \texttt{pRT}: \textit{equilibrium} & $4.90^{+14.16}_{-3.20}$ & $0.36^{+0.18}_{-0.16}$ &  $1.678\pm0.008$ & $2.90^{+0.05}_{-0.07}$ & $1076^{+181}_{-103}$ & $-0.74^{+1.80}_{-1.84}$ & $49\pm22$ & 1804.6 & 1.23 & 233\\

     \hline 
     \textbf{\jedi{} $R$=100 } & & & & & & & \\
     \poseidon{}: \textit{free - Model I (SO)} & - & - & $1.612^{+0.004}_{-0.006}$ &  $2.85^{+0.02}_{-0.02}$ &  $694^{+119}_{-79}$ & - &  $64^{+22}_{-23}$  & 479.25 & 1.69 & 48 \\
     \poseidon{}: \textit{ free - Model II } & - & - & $1.600^{+0.010}_{-0.020}$ &  $2.86^{+0.02}_{-0.02}$ &  $880^{+158}_{-126}$ &  $-1.45^{+2.21}_{-1.38}$ &  $54^{+27}_{-23}$ & 479.25 & 1.69 & 48 \\
     \poseidon{}: \textit{free - Model II ($\mu < 3$) } & - & - & $1.580^{+0.020}_{-0.040}$ &  $2.86^{+0.02}_{-0.02}$ &  $1004^{+367}_{-206}$ &  $-2.38^{+2.18}_{-1.17}$ &  $61^{+24}_{-24}$ & 478.47 & 1.70 & 48 \\
     \poseidon{}: \textit{equilibrium} & $6.31^{+15.49}_{-2.95}$ & $0.48^{+0.22}_{-0.19}$ & $1.57^{+0.02}_{-0.02}$ & $2.86^{+0.02}_{-0.02}$ & $1119^{+215}_{-128}$ & $-2.77^{+1.61}_{-0.83}$ & $64^{+21}_{-22}$ & 482.17 & 1.52 & 54   \\
     \\
    \texttt{pRT}: \textit{free - Model I (SO)} & - & - &
    $1.695\pm0.008$ &  $2.82\pm0.08$ &  $682^{+184}_{-111}$ & $-0.02^{+1.31}_{-1.29}$ &  $57\pm22$  & 476.0 & 1.37 & 49 \\
     \texttt{pRT}: \textit{ hybrid} & $3.72^{+11.77}_{-2.43}$ & $0.34^{+0.18}_{-0.15}$ &
     $1.680\pm0.007$ &  $2.87^{+0.06}_{-0.07}$ &  $1016^{+137}_{-82}$ & $-2.14^{+2.41}_{-0.73}$ &  $50\pm21$  & 476.3 & 1.39 & 51 \\
     \texttt{pRT}: \textit{equilibrium} & $3.63^{+8.67}_{-2.08}$ & $0.37^{+0.17}_{-0.17}$ &  $1.680\pm0.007$ & $2.88^{+0.05}_{-0.06}$ & $1063
     ^{+137}_{-82}$ & $-0.64^{+1.79}_{-1.68}$ & $42\pm22$ & 477.5 & 1.40 & 54\\

     \hline 
     \textbf{\tiberius{} $R$=400 } & & & & & & & \\
     \poseidon{}: \textit{free - Model I (SO)} & - & - & $1.610^{+0.010}_{-0.010}$ &  $2.86^{+0.02}_{-0.02}$ &  $723^{+138}_{-96}$ & - & $48^{+22}_{-22}$  & 1814.79 & 1.25 & 228  \\
     \poseidon{}: \textit{ free - Model II } & - & - & $1.590^{+0.010}_{-0.020}$ &  $2.86^{+0.02}_{-0.02}$ &  $935^{+128}_{-137}$ &  $-2.51^{+1.95}_{-0.72}$ &  $44^{+20}_{-19}$ & 1813.92 & 1.26  & 228  \\
     \poseidon{}: \textit{free - Model II ($\mu < 3$) } & - & - & $1.580^{+0.020}_{-0.020}$ &  $2.86^{+0.02}_{-0.02}$ &  $964^{+146}_{-141}$ &  $-2.70^{+0.84}_{-0.65}$ &  $46^{+20}_{-20}$ & 1813.48 & 1.26 & 228  \\
     \poseidon{}: \textit{equilibrium} & $3.39^{+8.51}_{-1.70}$ & $0.37^{+0.20}_{-0.12}$ & $1.57^{+0.01}_{-0.02}$ & $2.86^{+0.02}_{-0.02}$ & $1058^{+175}_{-80}$ & $-2.47^{+1.07}_{-0.83}$ & $62^{+21}_{-22}$ & 1815.85 & 1.24 & 234\\
     \\
    \texttt{pRT}: \textit{free - Model I (SO)} & - & - &
     $1.695\pm0.006$ &  $2.83^{+0.05}_{-0.06}$ &  $602^{+110}_{-67}$ & $0.11^{+1.26}_{-1.27}$ & $56\pm20$ &  1809.5 & 1.29 & 229 \\
     \texttt{pRT}: \textit{ free - Model II } & - & - &
     $1.679\pm0.010$ &  $2.88^{+0.06}_{-0.07}$ &  $914^{+174}_{-145}$ & $-2.43^{+1.11}_{-0.72}$ & $57\pm20$ &  1807.5 & 1.28 & 230 \\
     \texttt{pRT}: \textit{ hybrid} & $2.04^{+5.90}_{-1.25}$ & $0.30^{+0.19}_{-0.13}$ &
     $1.682\pm0.006$ &  $2.87^{+0.06}_{-0.07}$ &  $987^{+113}_{-71}$ & $-2.26^{+2.12}_{-0.63}$ & $59\pm22$ &  1809.7 & 1.27 & 231 \\
     \texttt{pRT}: \textit{equilibrium} & $1.70^{+4.33}_{-1.01}$ & $0.29^{+0.17}_{-0.13}$ & $1.683 \pm 0.006$ & $2.87^{+0.06}_{-0.07}$ & $997^{+124}_{-71}$ & $-1.26^{+2.18}_{-1.44}$ & $49\pm24$ &  1810.4 & 1.28 & 234\\

     \hline 
     \textbf{\tiberius{}{} $R$=100 } & & & & & & & \\
     \poseidon{}: \textit{free - Model I (SO)} & - & - & $1.612^{+0.004}_{-0.005}$ &  $2.85^{+0.03}_{-0.02}$ &  $650^{+102}_{-73}$ & - & $62^{+23}_{-23}$ &  492.44 & 1.20 & 48 \\
     \poseidon{}: \textit{ free - Model II } & - & - & $1.600^{+0.010}_{-0.010}$ &  $2.86^{+0.02}_{-0.02}$ &  $813^{+145}_{-113}$ &  $-1.18^{+1.94}_{-1.30}$ &  $46^{+21}_{-20}$ & 489.11 & 1.37  & 48  \\
     \poseidon{}: \textit{free - Model II ($\mu < 3$) } & - & - & $1.590^{+0.010}_{-0.030}$ &  $2.86^{+0.02}_{-0.02}$ &  $850^{+250}_{-147}$ &  $-1.80^{+2.23}_{-1.29}$ &  $49^{+21}_{-21}$ & 488.46 & 1.38 & 48 \\
     \poseidon{}: \textit{equilibrium} & $3.63^{+9.55}_{-1.70}$ & $0.38^{+0.21}_{-0.13}$ & $1.57^{+0.01}_{-0.02}$ & $2.86^{+0.02}_{-0.02}$ & $1085^{+204}_{-98}$ & $-2.57^{+1.66}_{-0.91}$ & $61^{+22}_{-23}$ & 490.80 & 1.24 & 54 \\
     \\
    \texttt{pRT}: \textit{free - Model I (SO)} & - & - &
    $1.695\pm0.006$ &  $2.83\pm0.06$ &  $610^{+113}_{-70}$ &  $0.13^{+1.18}_{-1.21}$ &  $63\pm21$ & 488.0 & 1.18 & 49 \\
     \texttt{pRT}: \textit{ hybrid} & $2.14^{+5.81}_{-1.40}$ & $0.31^{+0.19}_{-0.14}$ &
     $1.680\pm0.008$ &  $2.86^{+0.06}_{-0.08}$ &  $1014^{+159}_{-90}$ &  $-2.46^{+1.98}_{-0.70}$ &  $59\pm23$ & 487.7 & 1.22 & 51 \\
     \texttt{pRT}: \textit{equilibrium} & $1.82^{+3.55}_{-1.10}$ & $0.31^{+0.18}_{-0.14}$ & $1.680\pm0.007$ &  $2.88^{+0.06}_{-0.07}$ & $1022^{+141}_{-82}$ &  $-1.30^{+2.17}_{-1.51}$ &  $48\pm23$ & 488.6 & 1.27 & 54\\

    \end{tabular}

\end{table*}
\setlength{\tabcolsep}{6pt}

\setlength{\tabcolsep}{6pt}

\section{Retrieval corner plots}\label{sec:cornerplots}

\begin{figure*}
    \centering
    \includegraphics[width=\linewidth]{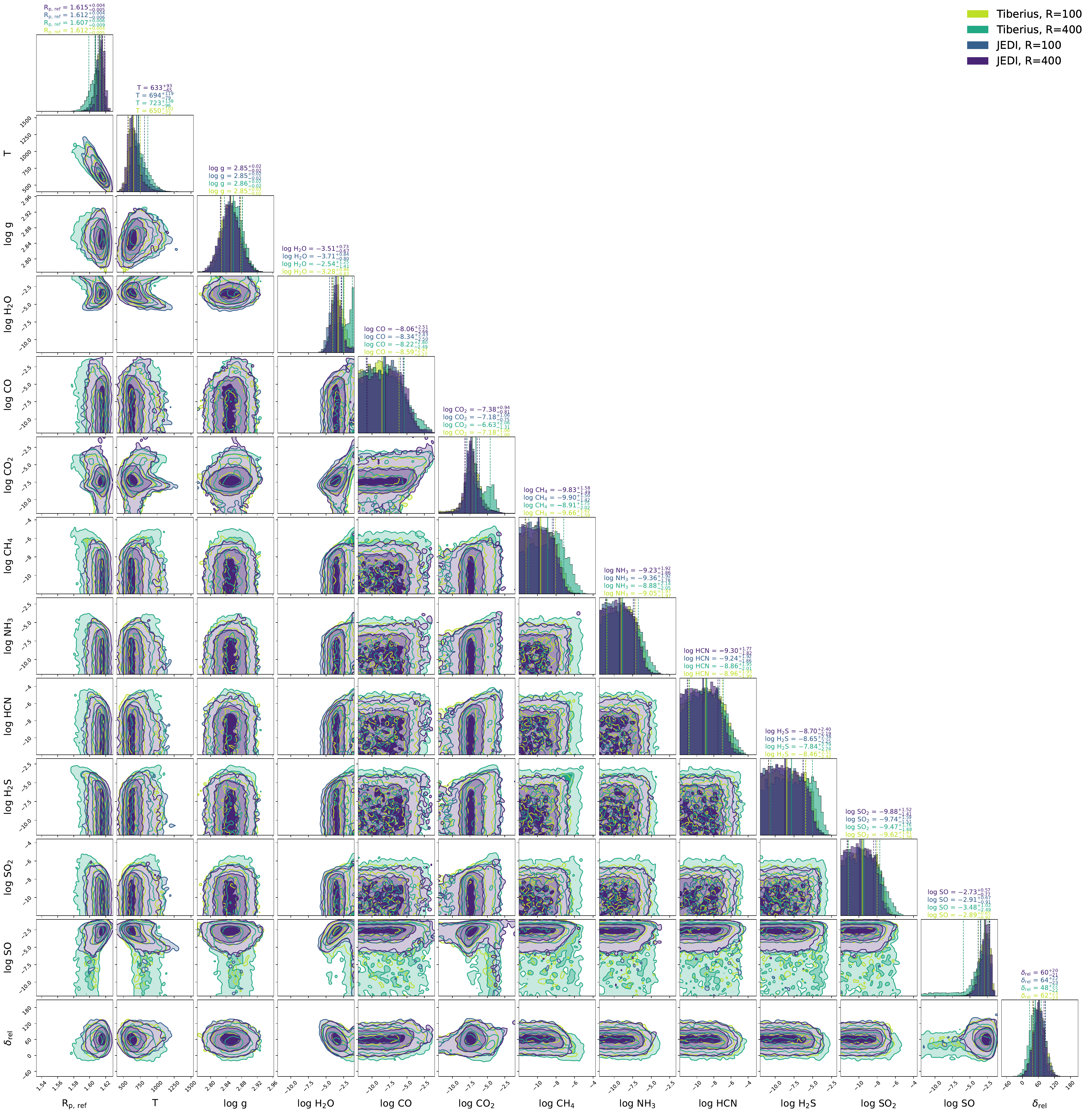}
    \caption{Retrieved posterior distribution for the POSEIDON free chemistry retrievals for model I (including \ce{SO} opacity). The species abundances are parameterised as volume mixing ratios.}
    \label{fig:poseidon_spec_hist_m1}
\end{figure*}

\begin{figure*}
    \centering
    \includegraphics[width=\linewidth]{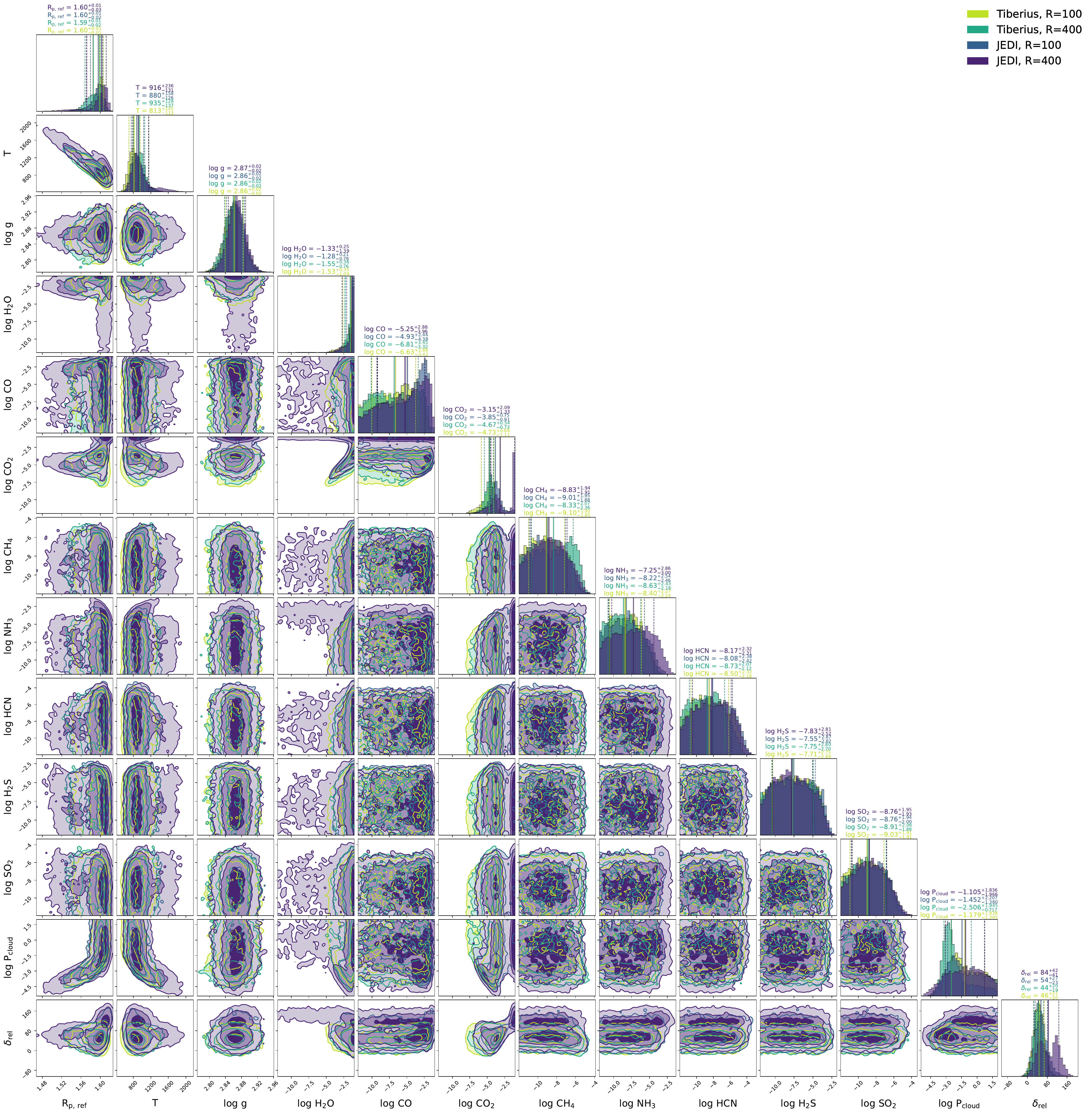}
    \caption{Retrieved posterior distribution for the POSEIDON free chemistry retrievals for model II. The species abundances are parameterised as volume mixing ratios.}
    \label{fig:poseidon_spec_hist_m2}
\end{figure*}

\begin{figure*}
    \centering
    \includegraphics[width=\linewidth]{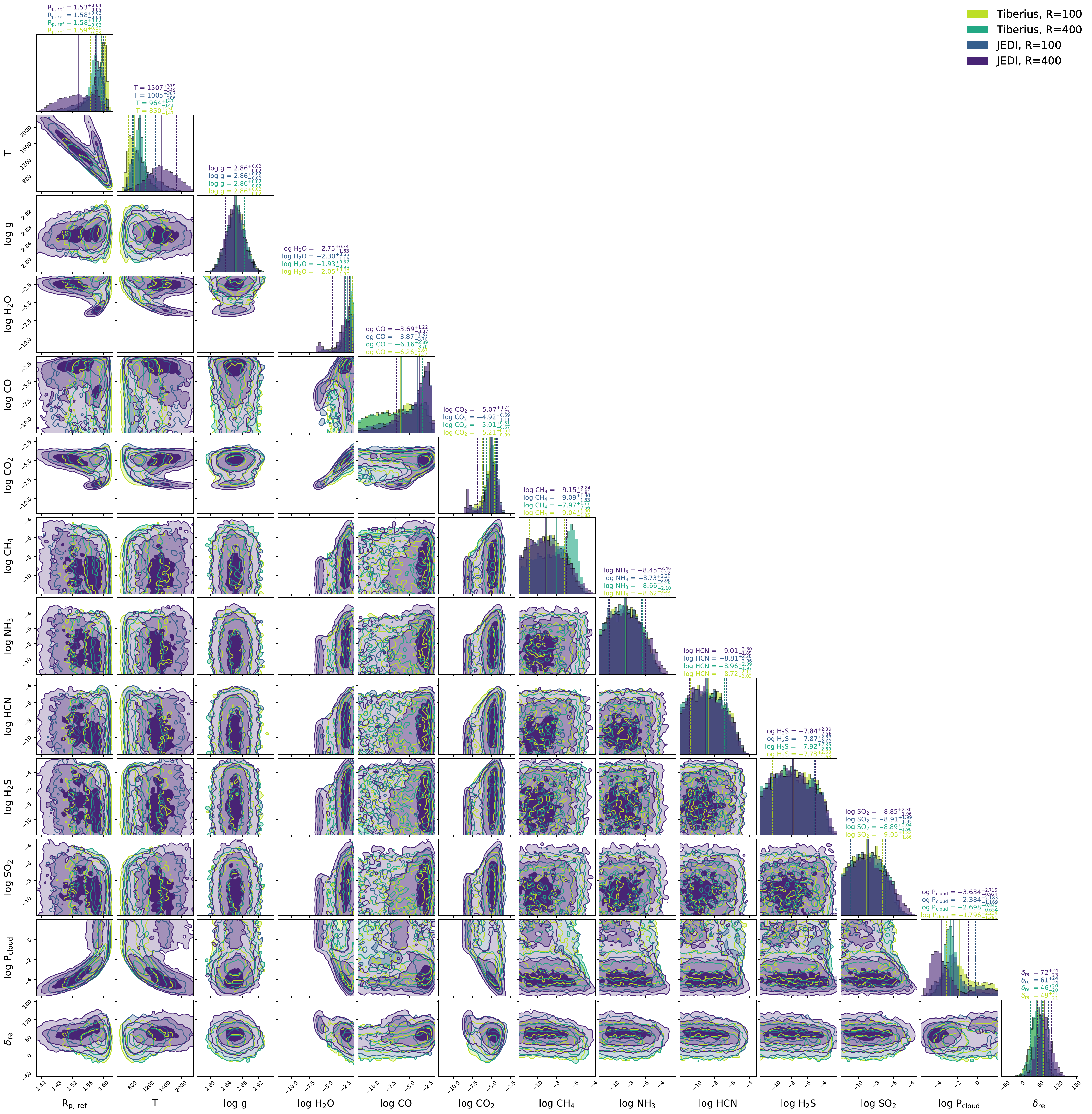}
    \caption{Retrieved posterior distribution for the POSEIDON free chemistry retrievals for model II ($\mu<3$). The species abundances are parameterised as volume mixing ratios.}
    \label{fig:poseidon_spec_hist_m3}
\end{figure*}

\begin{figure*}
    \centering
    \includegraphics[width=\linewidth]{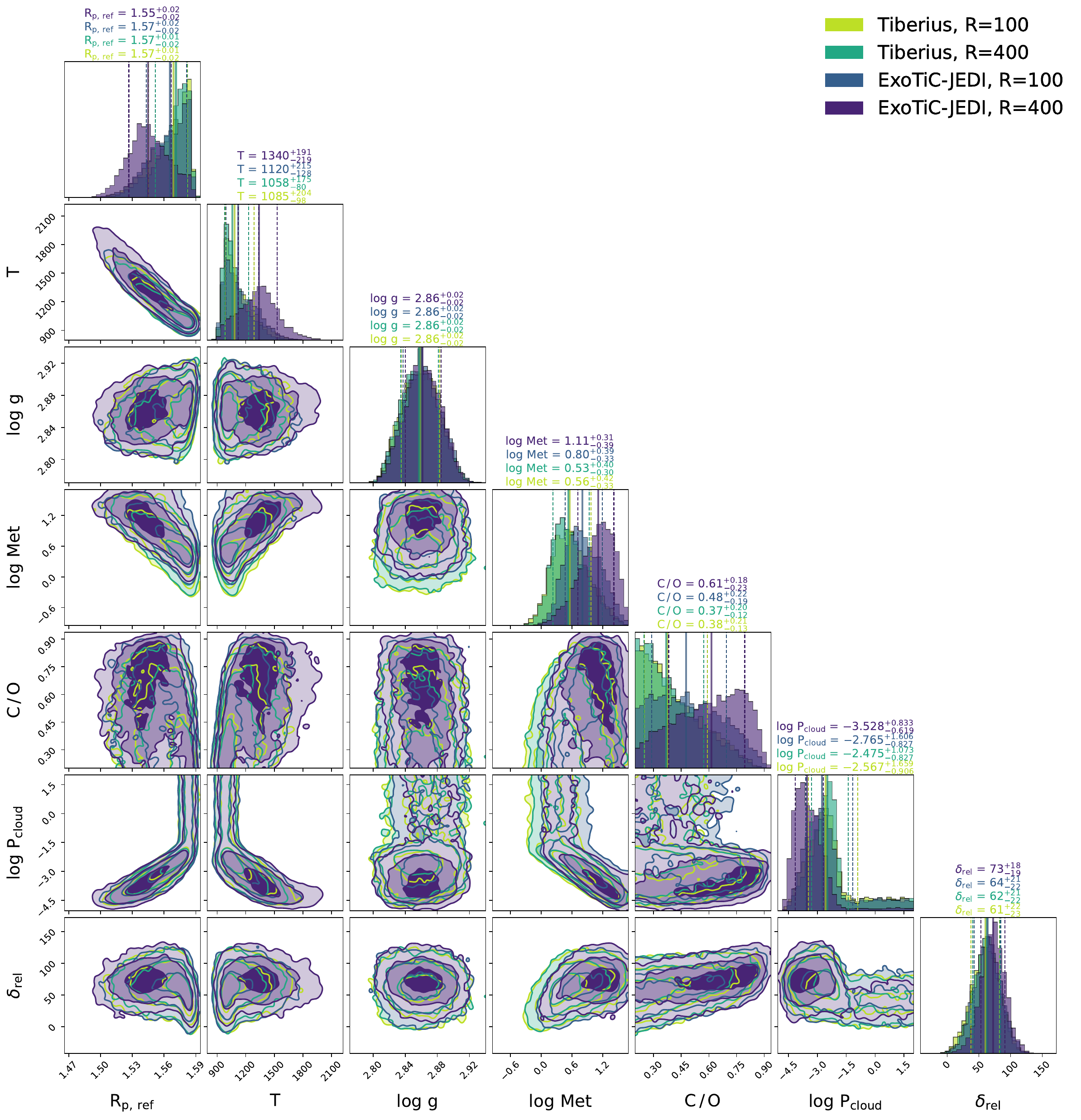}
    \caption{Retrieved posterior distribution for the POSEIDON equilibrium chemistry retrievals with an upper metallicity prior limit of 50x solar ($[\mathrm{M/H}]$ = 1.7).  }
    \label{fig:poseidon_spec_hist_eq}
\end{figure*}

\begin{figure*}
    \centering
    \includegraphics[width=\linewidth]{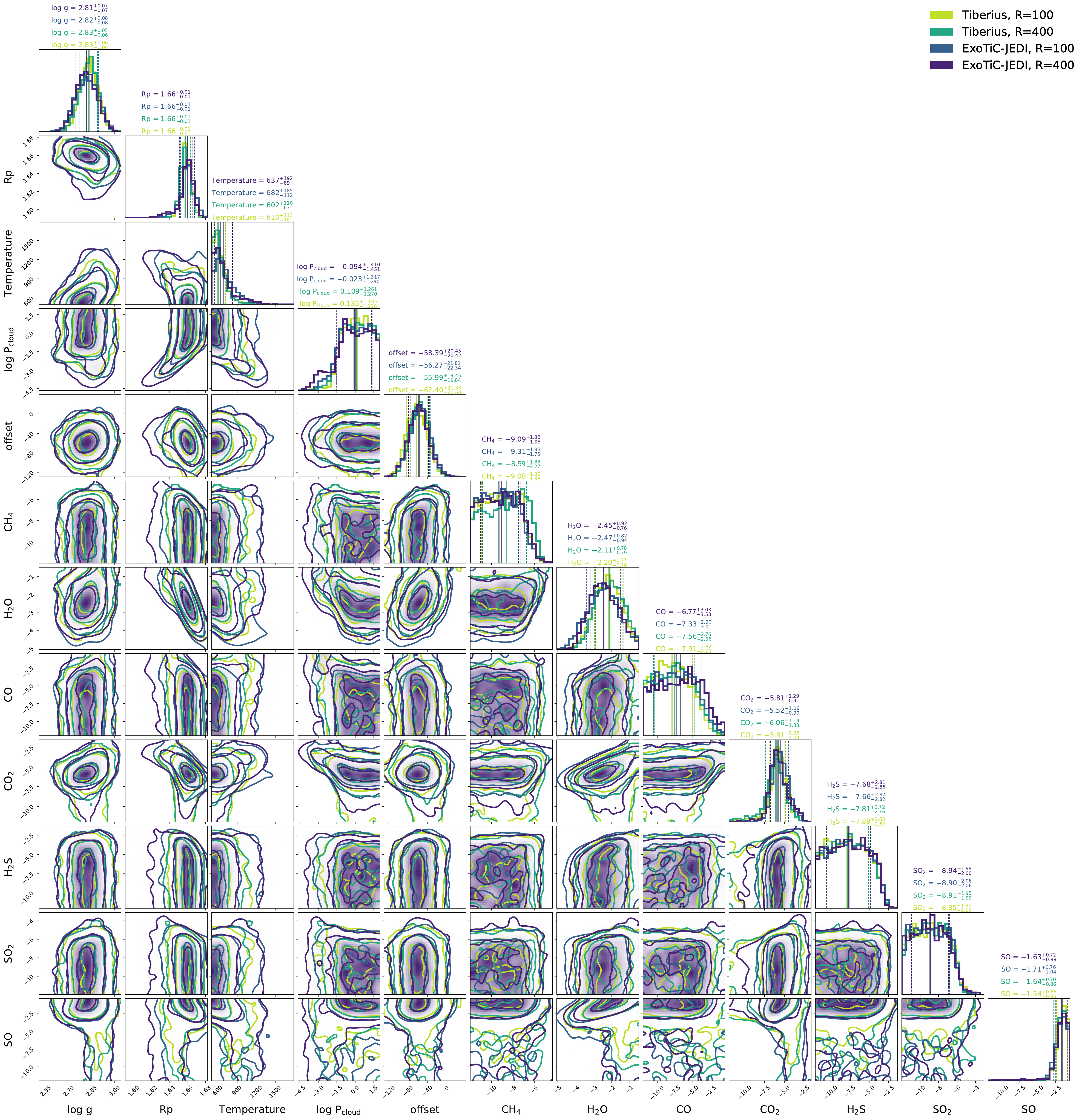}
    \caption{Retrieved posterior distribution for the \prt{} free chemistry retrievals for model I (including \ce{SO} opacity). The species abundances are parameterised as mass fractions. }
    \label{fig:prt_overplot_model_i}
\end{figure*}

\begin{figure*}
    \centering
    \includegraphics[width=\linewidth]{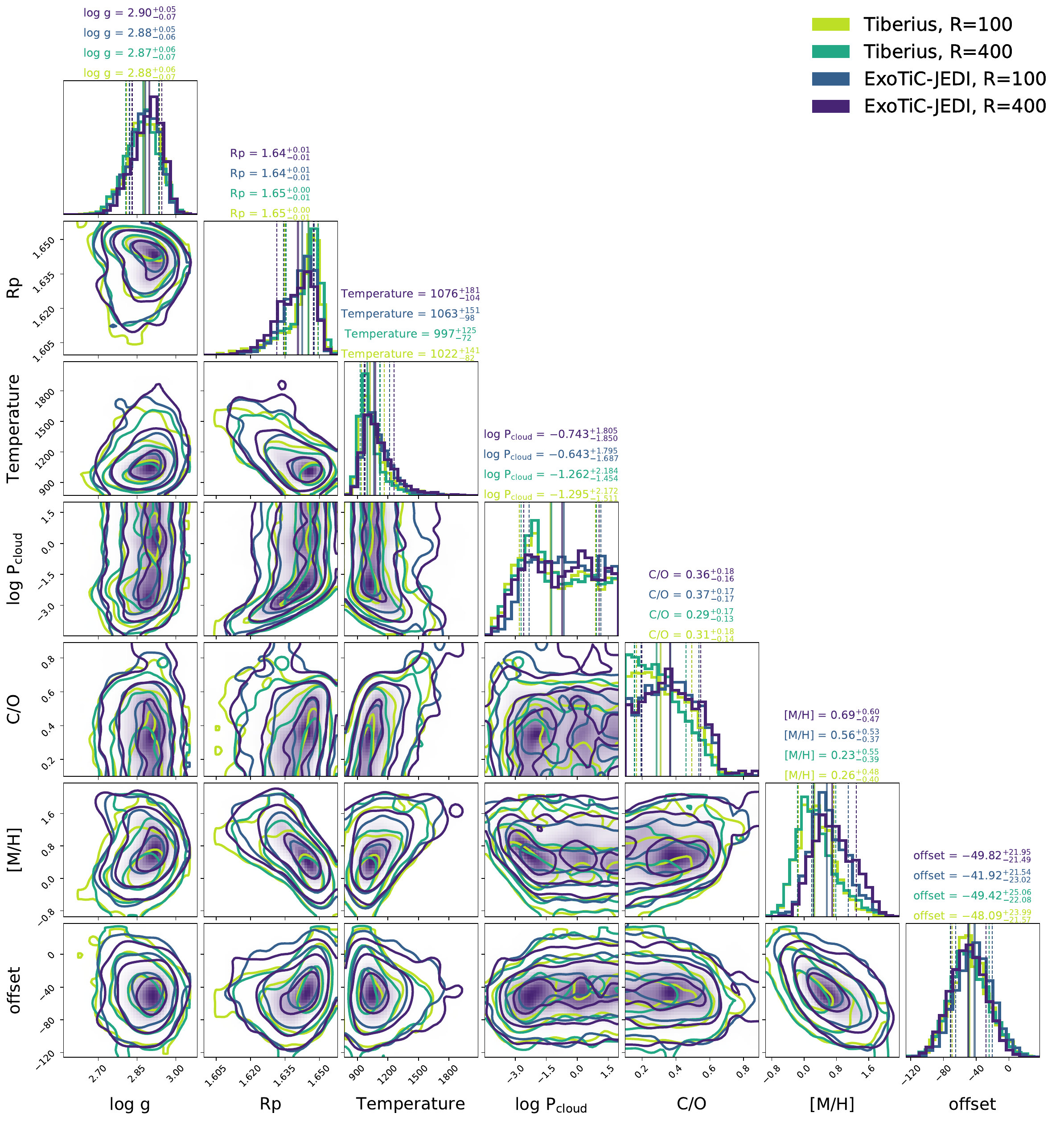}
    \caption{Retrieved posterior distribution for the \prt{} equilibrium chemistry retrievals. }
    \label{fig:prt_overplot_equilibrium}
\end{figure*}

\bsp	
\label{lastpage}
\end{document}